# Steps towards a theory and calculus of aliasing

**Bertrand Meyer**
**ETH Zurich & Eiffel Software**
**se.ethz.ch**

*Abstract*

A theory, graphical notation, mathematical calculus and implementation for finding whether two given expressions can, at execution time, denote references attached to the same object. Intended as the seed for a comprehensive solution to the "frame problem" and as an alternative (for the specific issue of determining aliases) to separation logic, shape analysis, ownership types and dynamic frames[1].

## 1 Dynamic aliasing

You have, most certainly, read Homer. I have not (too much blood), but then I listen to Offenbach a lot, so we share some knowledge: we both understand that "*the beautiful daughter of Leda and the swan*", "*poor Menelaus's spouse*" and "*Pâris's lover*" all denote the same person, also known as *Helen of Troy*. The many modes of calling Helen are a case of *aliasing*, the human ability to denote a single object by more than one name.

Aliasing is at its highest risk of causing confusion when it is *dynamic*, that is to say when an object can at any moment acquire a new name, especially if that name previously denoted another object. The statement "*I found Pâris's lover poorly dressed*" does not necessarily cast aspersion on Helen's sartorial tastes, as Pâris might by now have found himself a new lover; but if we do not carefully follow the lives of the rich and famous we might believe it does.

Stories of dire consequences of dynamic aliasing abound in life, literature and drama. There is even an opera, Smetana's *The Bride Sold*[2], whose plot *entirely* rests on a single aliasing event. To the villagers' dismay, Jeník promises the marriage broker, in return for good money, not to dissuade his sweetheart Mařenka from marrying the son of the farmer Mícha. Indeed Mícha wants Mařenka for his dimwit son, Vašek, but it is suddenly revealed that Jeník, believed until then to be a stranger to the village, is Mícha's son from a first marriage: he has tricked everyone.

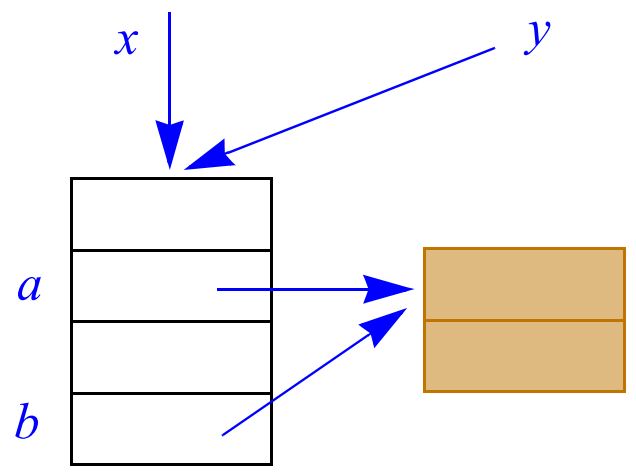

To a programmer, this tale sounds familiar: the equivalent in program execution is to perform an operation on certain operands, and inadvertently to modify a property of a target that is not named in the operation — hence the risk of confusion — but aliased to one of the operands. For example an operation may, officially, modify the value of $x.a$; but if $x$ denotes a reference and $y$ another reference which happens at the time of execution to be aliased to $x$ (meaning that they both point to the same object), the operation will have an effect on $y.a$ even though its text does not cite $y$. If $b$ is aliased to $a$, we might even have an operation that modifies $y.b$ even though its description in the programming language mentions neither $y$ nor $b$.

1. This paper is a revision of "Towards a Theory and Calculus of Aliasing", *Journal of Object Technology*, Vol. 9, no. 2, March 2010, pages 37-74, www.jot.fm/issues/issue_2010_03/column5.pdf. Along with numerous corrections, it brings a proper treatment of routine arguments and an improved theoretical basis.
2. A title incorrectly rendered, in the standard English translation, as *The Bartered Bride*.

It is not hard to justify the continued search for effective verification techniques covering aliasing. In the current state of proof technology, the aliasing problem (together with the closely related *frame* problem, to which it provides the key) is the principal obstacle preventing full proofs of correctness for sequential programs. It also plays a role in the specific difficulties of proving *concurrent* programs correct. (For references on the issues and approaches cited in this section, see appendix F.)

A symptom of this situation is that industrial program proving tools often preclude the use of pointers altogether. An example is the Spark environment, which has made a remarkable contribution towards showing that production programs can be routinely subjected to proof requirements. Spark relies on a programming language, presented as a subset of Ada but in reality a subset of a Pascal-like language (plus modules), without support for pointers or references. In considering how to make such pioneering advances relevant to a larger part of the industry, it is hard to imagine masses of programmers renouncing pointers and other programming languages advances of the past three decades.

The absence of a generally accepted solution is not due to lack of trying. The aliasing problem has been extensively researched, and interesting solutions proposed, in particular *shape analysis*, *separation logic*, *ownership types* and *dynamic frames*. Few widely used proof environments have integrated these techniques. That may still happen, but the obstacles are significant; in particular, the first two approaches suffer (in our opinion) by attempting to draw a picture of the run-time pointer structure that is more precise than needed for alias analysis; and the last three assume a supplementary annotation effort (in addition to standard Hoare assertions) at which programmers may balk.

The theory, calculus and prototype implementation described here strive to avoid these limitations. A representative application is to prove the absence of aliasing between any elements of two linked lists created and modified through typical object-oriented techniques. Assume a standard implementation of lists with an operation to add elements at the end:

```
In class LIST:
   extend (a: ELEMENT)
            -- Add cell at end, with a as item.
      local
            new, last: CELL
      do
                  -- First create new cell:
            create new ; new.set (a)
                  -- Then insert new cell at end:
            if first = Void then
                        -- List was empty. It will now consist of a single cell, the new one:
                  first := new
            else
                        -- List was not empty; go to its last cell and make last denote it:
                  from last := first until last.right = Void loop last := last.right end
                        -- Then link last cell to new cell:
                  last.set_right (new)
            end
      end
```

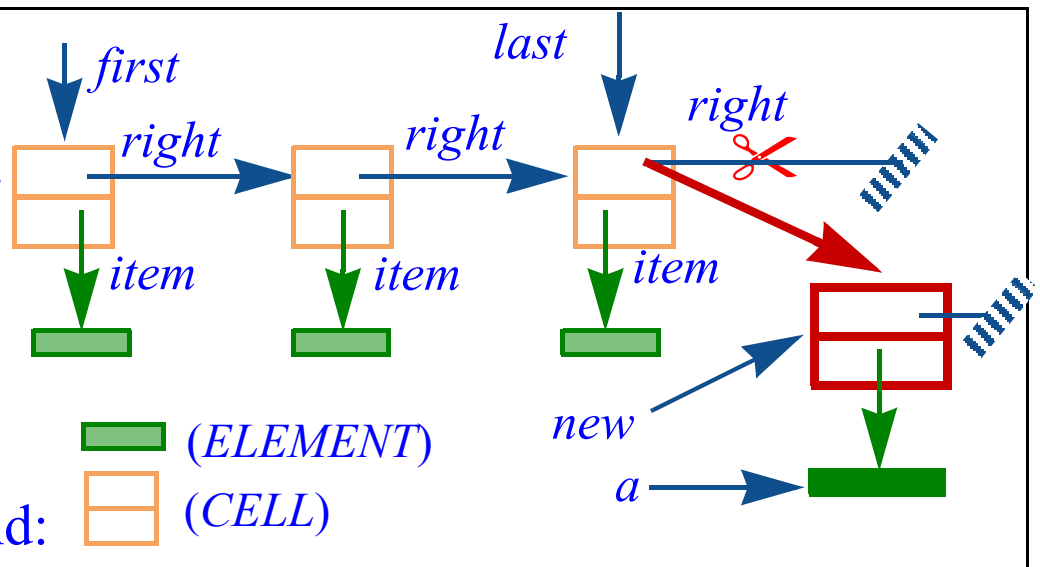

> With, in class *CELL*:
> *item*: *ELEMENT* ; *right: CELL*
> *set* (*v*: *ELEMENT*) **do** *item* := *v* **end**
> *set_right* (*c*: *CELL*) **do** *right* := *c* **end**

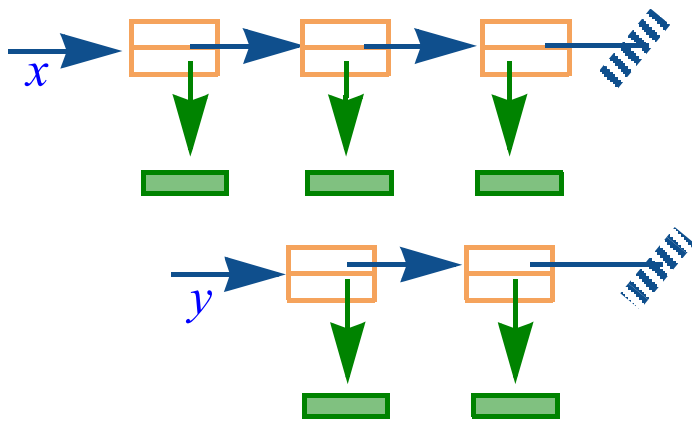

Consider references $x$ and $y$ denoting two such lists built through any number of applications of *extend* and similar operations. The theory, and its implementation, determine that if $x \neq y$ (the references are not aliased to each other) then no *CELL* or *ELEMENT* reachable from $x$ is also reachable from $y$. The proof is entirely automatic: it does not require any annotation. In the implementation, it is instantaneous.

In its present state the theory suffers from some limitations (section 8), but it makes the following claims:

- It provides a comprehensive treatment of aliasing issues and some progress towards a solution of the "frame problem".
- It includes a graphical notation, *alias diagrams*, which helps reason about aliasing.
- Alias analysis generally requires no assertions or other annotations. The only exception is the occasional need to add a **cut** instruction (4.5) to inform the calculus with results obtained from other sources. Outside of this case, alias analysis enjoys the advantage often invoked in favor of model checking and abstract interpretation against annotation-based approaches to program proving: full automation.
- The loss of precision (inevitable because of the undecidability of aliasing in its general form) is usually acceptable, and, when not, can be addressed through **cut**.
- The theory is at a suitably high level of abstraction, avoiding explicit references to such implementation-oriented concepts as "stack" and "heap".
- The theory can model the full extent of a modern object-oriented language.
- The reader will, it is hoped, agree that it is simple (about a dozen rules) and provides insights into the nature of programming, especially object-oriented programming. An example is the final rule /40/, for qualified calls: ($a$ » **call** $x.r$) = ($x$ ▪ (($x$' ▪ $a$) » $\underline{r}$)), which concisely conveys the essence of the fundamental mechanism of O-O computation, capturing the notion of current object and the principle of relativity, both central to the O-O model.

The following ideas are believed to be new (although of course strongly influenced by previous work): the notion of alias calculus; alias diagrams (a simplification of "shape graphs"); the canonical form of alias relations; limiting analysis to expressions occurring in the program; using alias analysis as a preprocessing step for axiomatic-style proofs; **cut**; negative references; the handling of arguments, loops and conditionals.

The ambition behind the present work is that it will complement the methods listed earlier and, for the problem of determining aliases (which is only a part of their scope), possibly provide an alternative.

Section 2 sets the context. Section 3 introduces alias relations. Section 4 presents the calculus for a simple language without remote object access, which section 5 extends with procedures. Section 6 generalizes the language and the calculus to the target domain of interest: object-oriented programming. Section 7 presents the prototype implementation. Section 8 summarizes how to apply the calculus to an actual object-oriented programming language and lists the remaining problems. Appendix A gives a formal model, appendices B and C the proofs of key properties, appendix D a sketch of a possible "must alias" calculus, and appendix E a list of all the rules of the calculus. Appendices F and G add acknowledgments and references.

All the examples of this article can be tried out in the implementation, which the reader can download (as a Windows executable) from se.ethz.ch/~meyer/down/alias.zip.

# 2 General observations

The goal of the calculus is to allow deciding whether two reference expressions appearing in a program might, during some execution, have the same value, meaning that the associated references are attached to the same object.

## 2.1 Adding the alias calculus to an axiomatic framework

To put the rest of the discussion in context, it is useful to describe where the approach may fit in the overall process of software verification. The key to the simplicity of the calculus is the expectation that aliasing is, in practice, the exception: most of the time, two expressions are *not* aliased to each other. As a consequence, the envisioned verification process is an incremental modification of standard axiomatic (Hoare-style) techniques:

A1 A first step uses the alias calculus to determine the possible aliases of expressions that appear in assertions. As a result, these expressions are enriched: for any assertion of the form *some_property* (*a*), this step adds *some_property* (*b*) for all *b* that can be aliased to *a*.

A2 The second step applies standard axiomatic reasoning to the program equipped with the resulting set of assertions — the original enriched with alias variants.

The techniques used in these two steps are independent. Step 2 uses ordinary axiomatic semantics (including backward reasoning because of the assignment axiom); step 1 uses the calculus (which happens to work in a forward style).

The following example illustrates the process. Assume we are asked to prove

```
{not y.a}                                   /1/
x.set_a
{not y.a}
```

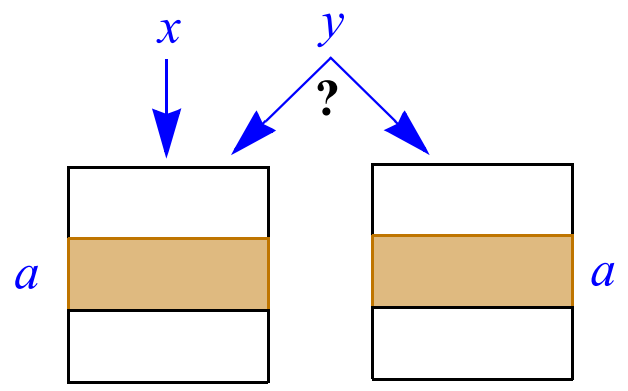

We are dealing with objects having a boolean attribute *a*, which the procedure *set_a* sets to **True**. Assume that we have at our disposal a proof framework which applies the standard techniques of axiomatic semantics, enabling us to prove

```
{True}                                      /2/
x.set_a
{x.a}
```

The proof of /2/ will involve the assignment axiom, as *set_a* performs *a* := **True**, and a procedure rule. (If we informally understand the call as *x.a* := **True**, the proof is a trivial application of the assignment axiom.)

If we naïvely applied similar techniques to prove /1/, the proof would proceed smoothly: since the instruction does not name *y*, the postcondition sails through that instruction unchanged. Such reasoning, however, is not sound if *y* can be aliased to *x*. The alias calculus will allow us, through its own techniques distinct from axiomatic semantics, to determine possible aliasings. If it finds that some computations might alias *y* to *x*, it will inform the axiomatic reasoning by automatically enriching the postcondition of /1/ to read **not** *y.a* **and not** *x.a*. Then /1/ is no longer a correct Hoare triple since application of the assignment axiom to **not** *x.a* yields the weakest precondition **False**.

## 2.2 Handling imprecision

For clarity of presentation it is useful to introduce the theory and calculus in terms of three successive programming languages of increasing ambition, each a superset of the previous one: E0 introduces variables and basic instructions; E1 introduces procedures; E2 introduces object-oriented mechanisms and represents the common core of modern O-O languages. To apply the calculus in practice, it will be necessary to translate the actual programming language of interest into E2. Elements of the following discussion, and the summary in section 8, describe how to perform the translation and, as a consequence, how to apply the calculus to practical programs written in such languages as Eiffel, Java or C#.

Until then, we will concentrate on the calculus itself. We must, however, note the principal property of the translation: it must be **sound**, meaning that if two expressions in the original language may become aliased in some execution the calculus must reflect that property. In the reverse direction, there is no such exigency: the calculus might tell us that two translated expressions can become aliased where the original expressions cannot. We may call this phenomenon *imprecision*. The presentation of the calculus will note cases in which the risk of imprecision arises.

This risk is inevitable in any practical approach to alias analysis, but might prevent some program proofs because of the possible loss of information. The theory introduces a special solution to this problem in the form of the **cut** instruction (section 4.5). A **cut** corrects any undesired imprecision resulting from the simplifications of the alias calculus by asserting at a particular point of the program that two expressions are *not* aliased. The alias calculus itself is not, in such cases, able to prove this property; the proof falls back on its partner in the proof duo — axiomatic semantics (step A2 above). As an example, consider

```
if not cond then
    x := y                                                        /3/
end
Other_instructions        -- Not affecting any of cond, x and y.
if cond then                                                      /4/
    z := x                                                        /5/
end
```

The alias theory correctly determines that at the start of the second conditional instruction /4/ *x* may be aliased to *y* as a result of the first one /3/. It will also determine, as a consequence, that the assignment /5/ may alias *z*, through *x*, to *y*. Such aliasing cannot occur in practice because the boolean expression *cond* has the same value in both cases. The alias calculus, however, has no way of establishing that no run-time execution path can include both /3/ and /5/; such a property is beyond its scope. If the imprecision is unacceptable — in other words, if the spurious aliasing of *z* to *y* precludes proving the properties of interest — we need to add a **cut** instruction to the second conditional, which becomes

```
if cond then
    cut x, y
    z := x
end
```

For the alias calculus, the **cut** instruction is a guarantee from the environment (as provided by **require** in Eiffel and **assume** in JML and Spec#) that *x* /= *y*. For the axiomatic proof framework, it is a proof obligation (**check** in Eiffel, **assert** in JML and Spec#).

## 2.3 Scope of the theory

The purpose of the alias theory and calculus is to answer a specific question:

> **The aliasing question**
> Given two expressions of a program, *e* and *f*, of reference type, and a program point *p*, can *e* and *f* ever be attached to the same object when an execution of the program is at *p*?

In line with the preceding observations, the calculus looks for a sound but possibly imprecise answer: it *may* — as rarely as possible — answer "yes" even if *e* and *f* could never become aliased in actual executions; but if they can, the calculus is *required* to answer "yes".

The most important word of the above definition is the first one: "*Given*". What makes the calculus possible is that it takes the pragmatic view of an existing program, possibly equipped with assertions. Then program proofs do not need to know all aliasing properties of all possible expressions; they only need the properties of *expressions actually appearing in the program and its assertions*. Expressions not named in the program are no more interesting to the prover than (except to the philosopher) the tree that falls unheard and unseen in the forest.

This observation allows us, in addition, to consider finite sets only. Without it, the analysis of a typical data structure traversal loop such as

```
from
    x := first
until some_condition loop
    x := x.right
end
```

would have to reflect that *x* can become aliased to *first*, *first.right*, *first.right.right* and so on, an infinite set of expressions. It might even force us to extend the assertion language with a regular-expression notation (*first.right**) to cover all possible values. While the alias calculus could accommodate such extensions, it does not need them for the fundamental applications discussed here.

The soundness requirement implies that not all the aliasings predicted by the calculus will necessarily arise during a particular execution. It could be useful to determine the aliasings that always happen ("*must alias*" rather than "*may alias*" relations). Such a study is beyond the present article, but appendix D gives some basic directions.

# 3 Alias relations

The theory relies on a notion of "alias relation", describing the possible aliasings between variables and expressions of a program.

## 3.1 Definition

**Definition: alias relation**

A relation in $E \leftrightarrow E$ for some set $E$ is an alias relation if it is symmetric and irreflexive.

$E \leftrightarrow E$, defined as $\mathbb{P}(E \times E)$, is the set of binary relations on $E$. For our needs $E$ will be a set of variables and expressions in a program. The presence of a pair $[x, y]$ in an alias relation associated with a program point expresses that $x$ and $y$ may be attached to the same object at that program point during some execution.

Such a relation must be symmetric: if $e$ may become aliased to $f$, then $f$ may become aliased to $e$. This possibility is only interesting in the case of different expressions, hence the requirement of irreflexivity ($e$ is not aliased to $e$).

If $r1$ and $r2$ are alias relations, so are $r1 \cup r2$, $r1 \cap r2$ and $r1 - r2$ (difference), but not, for example, the complement of $r1$.

If $r$ is a relation, but not necessarily an alias relation, $\overline{r}$ will denote the alias relation obtained from $r$ by removing all reflexive pairs and symmetrizing all pairs; for example $\overline{\{[x, x], [x, y], [y, z]\}}$ is the alias relation $\{[x, y], [y, x], [y, z], [z, y]\}$.

Formally, $\overline{r}$ is $(r \cup r^{-1}) — Id\,[E]$ where "—" is set difference and $Id\,[E]$ is the identity relation on $E$. If $r$ is an alias relation, then $\overline{r} = r$.

It is useful to extend this notation to an arbitrary subset $A$ of $E$, defining $\overline{A}$ as $\overline{A \times A}$. ($A \times A$ is the "universal" relation involving all pairs in $A$.) So $\overline{\{x, y, z\}}$ is $\{[x, y], [y, x], [x, z], [z, x], [y, z], [z, y]\}$.

For a set $A$ described by extension, as in this example, we may omit the braces, writing just $\overline{x, y, z}$. We may express any alias relation in a *union form* $\overline{T}, \overline{U}, \overline{V}, \ldots$, meaning $\overline{T} \cup \overline{U} \cup \overline{V} \ldots$, where every operand is a universal relation on a subset of $E$ with reflexive pairs removed. With this notation, we may write the first example, $\{[x, y], [y, x], [y, z], [z, y]\}$, as $\overline{x, y}$, $\overline{y, z}$.

An alias relation need not be transitive, as illustrated by the program extract

```
if cond then
    x := y
else
    x := z
    u := x
end
```

which, starting with no aliasing, yields (as the alias calculus will determine) the alias relation $\overline{x, y}$, $\overline{x, u, z}$ but does not cause aliasing between $y$ and $z$.

In dealing with alias relations, the following operator will be useful: $r \backslash\!\!- A$, where $r$ is a relation in $E \leftrightarrow E$ and $A$ is a subset of $E$, is $r$ deprived of any pair that involves a member of $A$ as first or second element. Formally: $r \backslash\!\!- A$ is $r — \overline{A \times E}$. If $r$ is an alias relation, so is $r \backslash\!\!- A$. (The operator's definition will be extended in 6.4 to cover dot expressions.)

We also define the **quotient** $a \,/\, x$ of an alias relation $a$ in $E \leftrightarrow E$ by an element $x$ of $E$ (similar to the equivalence class of $x$ in an equivalence relation) as the set containing $x$ and all the elements aliased to $x$:

$$a \,/\, x \;\triangleq\; \{y: E \mid (y = x) \vee [x, y] \in a\} \qquad \textbf{/6/}$$

## 3.2 Canonical form and alias diagrams

An alias relation may have several union forms; for example the union forms $\overline{x, y}, \; \overline{x, u, z}$ and $\overline{x, y}, \; \overline{x, u}, \; \overline{x, u, z}$ denote the same relation. The first of these variants, like all the examples given previously, is a **canonical form**:

**Canonical form of an alias relation**

The canonical form of an alias relation $a$ is a union form $T, U, V, \ldots$ where:

1. None of the sets $T, U, V, \ldots$ is a subset (proper or improper) of any of the others.
2. Adding or removing any element to or from any of them would invalidate the property $T \cup U \cup V \ldots = a$.

*Canonical form theorem*: For any alias relation $a$, the canonical form exists and is unique.

*Proof*: consider all subsets of $E$. Retain only those whose elements are all aliased to each other in $a$. Then remove any that is a subset of another. The resulting subset of $\mathbb{P}(E)$ gives a union form of $a$ that is canonical: any other subset $X$ of $\mathbb{P}(E)$ contains a subset of $E$ that either includes a non-aliased pair, so that $X$ violates condition 1 of the definition, or is a subset of an element of $X$, violating condition 2. ■

Although this is a constructive proof, an algorithm applying it directly to display the canonical form of a relation would be exponential in the size of $E$; the implementation uses a more efficient algorithm.

*Corollary*: each one of the sets $T, U, V, \ldots$ involved in a canonical form has at least two elements (since an alias reflection is irreflexive).

The reverse theorem also holds: a canonical form defines an alias relation uniquely. All alias relations for the examples that follow will be given in canonical form.

**Alias diagrams** are useful to visualize the theory and in particular the canonical form theorem. An alias diagram is a labeled directed graph with one special *source node* representing a program point and any number of *value nodes* each representing a set of possible values (not specified) in associated program states. At this stage of the theory, the graph is acyclic, the start node of any edge is the source node, and the end node is a value node; when we extend the theory to object-oriented programming in section 6, there will also be edges connecting value nodes. An example alias diagram is:

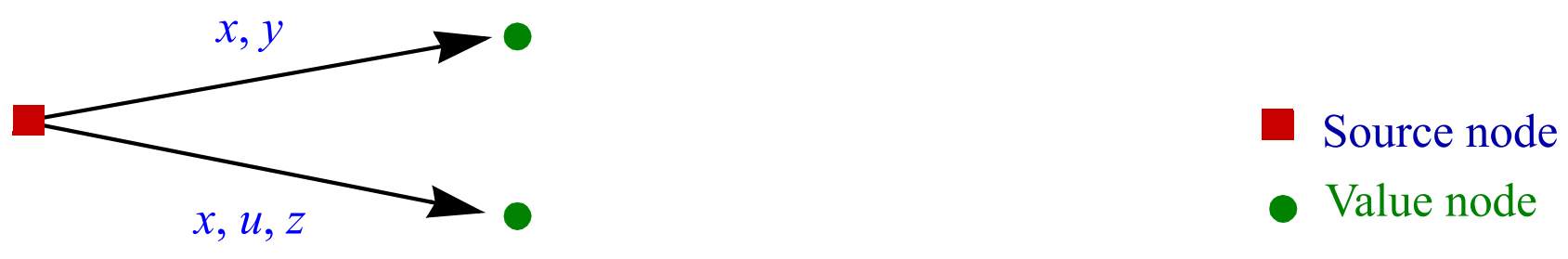

The label of every edge is a non-empty set of expressions, for example $e, f$. The presence of an expression $e$ in the label of an edge leading to a value node $n$ expresses that $e$ may at the given program point have one of the values associated with $n$. The alias relation associated with such a graph is simply the set of pairs $[e, f]$ such that $e$ and $f$ both appear in the label for some edge. So the graph above represents the earlier example $\overline{x, y}$, $\overline{x, u, z}$.

A value node carries no information other than its existence and the label of the edge (a single one at this stage of the theory) that leads to it. In the following discussion, as a consequence, "removing an edge" also implies removing the target node.

A diagram is in canonical form if no label is a subset of another. The canonical form theorem is easy to visualize on alias diagrams: a non-canonical diagram such as represents the same alias relation as the previous one, so the edge labeled $x, u$ is useless. To turn an arbitrary diagram into canonical form, remove any edge whose label is a subset of another edge's label (and, per the general convention, remove the edge's end node).

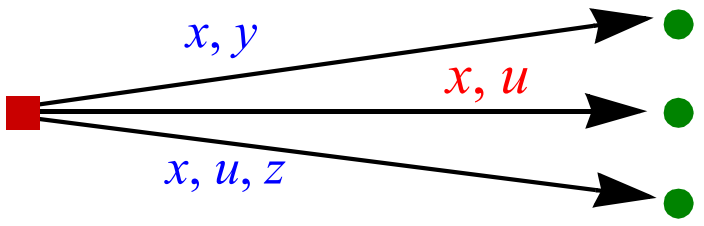

As a consequence of the corollary of the canonical form theorem, the label of every edge includes at least two expressions. One-expression labels 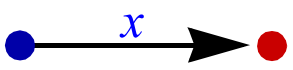  , expressing that $x$ may have a value at the current program point, may be interesting for other applications but are irrelevant for the theory of aliasing, at least until it gets extended for object-oriented programming.

### 3.3 The semantics of an alias relation

If $a$ is an alias relation on the set $E$ of reference variables and expressions appearing in a program $p$, we may associate with $a$ a Hoare-style assertion — a property of the program state written as an expression involving the values of program variables — written $a^{-}$ and defined as

$$a^{-} \triangleq \bigwedge_{[x,\, y] \notin a} x \neq y \quad \text{(where } x \text{ and } y \text{ are different variables)} \qquad /7/$$

where $\wedge$ is "and" between assertions. In words: $a^{-}$ states that if two distinct variables may ever have equal values, their pair appears in $a$; or equivalently, that pairs not appearing in $a$ are guaranteed to have different values. This notion reflects the conservative nature of the alias calculus: while the *presence* of a pair $[x, y]$ in $a$ states that $x$ *might* become aliased to $y$ but does not imply that it will, its *absence* from $a$ implies, for soundness, that $x$ *cannot* become aliased to $y$.

If the set of variables is given, the correspondence defined by /7/ between alias relations such as $a$ and assertions of the form appearing on the right for $a^{-}$ (a conjunction of clauses stating that variable pairs have different values) is one-to-one.

The $^{-}$ operator satisfies the following monotonicity theorem:

$$\text{If } a \subseteq b, \text{ then } a^{-} \Rightarrow b^{-} \qquad /8/$$

where $\Rightarrow$ is "implies" between assertions. (*Proof*: if $a^{-}$ holds, any pair $[x, y]$ that may have equal values must appear in $a$, and hence in $b$.)

## 3.4 Characterizing the effect of programs on aliasing

Aliasing is not compositional, in the naïve sense of allowing the definition of a function *aliases* such that *aliases* $(p)$ would determine the alias relation induced by the program $p$ in terms of *aliases* $(p_i)$ for components $p_i$ of $p$. Consider

| | |
|---|---|
| $p1$: | $x := y$ |
| $p2$: | $z := x \,;\, x := u$ |
| $p1 \,;\, p2$: | $x := y \,;\, z := x \,;\, x := u$ |

then *aliases* $(p1)$ would be $\overline{x, y}$, *aliases* $(p2)$ would be $\overline{x, u}$ and *aliases* $(p1; p2)$ would be the relation $\overline{y, z},\ \overline{x, u}$, which cannot be obtained by combination of the previous two since neither of them mentions $z$.

Instead, the calculus works on formulae of the form

$$a \gg p \qquad \textbf{/9/}$$

where $a$ is an alias relation and $p$ is a program component. /9/ denotes the alias relation that holds at the end of an execution of $p$ started in a state where $a$ held. In practice, both $a$ and $a \gg p$ may be conservative approximations of the actual alias relation. When considering an entire program $p$, we will be interested in $\varnothing \gg p$ for the empty relation $\varnothing$; the computation of $\varnothing \gg p$ will also yield the value of the alias relation at key program locations such as routine exit points.

A simplified interpretation of the meaning of $a \gg p$ in terms of Hoare semantics is

$$\{a^-\} \quad p \quad \{(a \gg p)^-\} \qquad \textbf{/10/}$$

stating that if we use $a$ as a guarantee about non-aliased pairs on entry to $p$, the calculus yields a guarantee about non-aliased pairs on exit.

/10/, sufficient for most practical purposes, is not a weakest precondition rule because it ignores the property that any aliasing in $a$ that affects a variable modified by $p$ is irrelevant, as it will be overridden by $p$. Consider for example a program with three variables $x$, $y$, $z$; take for $p$ the assignment $x := y$, and for $a$ the empty alias relation $\varnothing$. Then $a^-$ is the assertion stating that variables are pairwise different: $x \neq y$ **and** $x \neq z$ **and** $y \neq z$. It is intuitively clear, and given by the calculus (rules /17/ and /18/ below), that $a \gg p$ is $\overline{x, y}$. The corresponding assertion $(a \gg p)^-$ is $x \neq z$ **and** $y \neq z$. /10/ in this case correctly states that

$$\{x \neq y \textbf{ and } x \neq z \textbf{ and } y \neq z\} \quad x := y \quad \{x \neq z \textbf{ and } y \neq z\}$$

but the precondition is stronger than needed: its first two conditions, $x \neq y$ and $x \neq z$, are irrelevant since the assignment overwrites $x$. The weakest precondition in this case is just $y \neq z$ (as implied by the Hoare assignment rule). In the general case, the weakest precondition rule is (rather than /10/) the following:

**Alias calculus soundness rule**

For any relation $a$ and any construct $p$:

$$\{(a \cup \overline{(p\leftarrow \times E)})^-\} \quad p \quad \{(a \gg p)^-\} \qquad \textbf{/11/}$$

where $p\leftarrow$ is the set of variables that every terminating execution of $p$ sets.

$\overline{p\leftarrow \times E}$ is the set of pairs involving any of the variables modified by $p$; in the example, where $p\leftarrow$ is $\{x\}$, this set is $\overline{x, y}$, $\overline{y, z}$. The rule tells us that we may remove any of its pairs from $a$ in the precondition, since executing $p$ makes them irrelevant to the resulting state (as the rule for assignment, /17/ and /18/, will reflect).

Note that $p\leftarrow$ is the set of variables that *every* terminating execution of $p$ will set. If *some* executions of $p$, but not all, may set $x$, we cannot remove from $a$ the pairs involving $x$: for example if $a \gg p$ does not contain $\overline{x, z}$, meaning that the postcondition includes $x \neq z$, then this property must also appear in the precondition. Otherwise $x$ and $z$ could be initially aliased to each other, and the executions that do not set $x$ would leave them aliased on exit.

Obtaining determination of $p\leftarrow$ for all the constructs $p$ of any realistic programming language is undecidable; the proof of this result appears in appendix B. As a consequence, the weakest precondition rule /11/ is of theoretical use only. The appendix shows how obtain a reasonable under-approximation of $p\leftarrow$ through a simple syntactic rule, yielding a version of the soundness rule that is significantly more precise (in the sense of using a weaker precondition) than the simplified version, /10/. Using an under-approximation is sound since replacing $p\leftarrow$ by a subset means also replacing $a \cup \overline{(p\leftarrow \times E)})$ by a subset of this relation and hence, according to the monotonicity theorem /8/, replacing the associated assertion $(a \cup \overline{(p\leftarrow \times E)})^-$ by a stronger one.

This discussion only affects the theoretical connection between the alias calculus and axiomatic semantics, not the power of the calculus. But it has the practical consequence of confirming that the "forward" nature of the calculus is necessary rather than accidental. Standard Hoare-style and weakest-precondition semantics work "backward" because of the classic assignment axiom $\{P\,[y/x]\}\ x := y\ \{P\}$, which derives the precondition from the postcondition. In contrast, the rules of the alias calculus, as given below, derive $a \gg p$ from $a$, in forward style. To obtain a weakest precondition in /11/, we need $p\leftarrow$, which cannot be computed precisely in the general case. We may express this observation as a theorem:

> **Forward alias rule theorem**
>
> For the alias properties of any realistic language involving reference assignment, no weakest-precondition backward calculus is possible.

Even if the rules of axiomatic semantics give us a weakest precondition, it may not correspond to an alias relation. For example **(if** $c$ **then** $x := z$ **end) wp** $\overline{(x, y)}^-$ is the assertion **not** $c$ **and** $\overline{(x, y)}^-$, which for arbitrary $c$ is usually not of the form $a^-$ for an alias relation $a$.

The rules of the calculus will now follow, each defining $a \gg p$ for a given kind of instruction $p$. To be acceptable, each must guarantee that if $a$ is an alias relation so is $a \gg p$. In addition, every rule must satisfy the fundamental soundness rule /11/. The present discussion does not include a complete proof but gives an example, for one of the rules, in section 4.12.

## 4 The basic calculus

The first level of the calculus relies on a simple programming language, E0. The following subsections introduce the constructs of E0, their informal semantics, and the corresponding alias calculus rules.

In E0, all variables denote references; the value of a reference is an object identifier. The following presentation relies on an intuitive understanding of the instructions; a formal definition of E0 appears in A.1 (part of appendix A).

## 4.1 Skip

It is convenient to include a null instruction **skip** with the rule

$$a \text{ » } \mathbf{skip} \quad = \quad a \qquad \textbf{/12/}$$

( Shaded lines will signal rules of the alias calculus.)

## 4.2 Compound

If $p$ and $q$ are E0 instructions, the notation $p \,;\, q$ denotes an instruction that executes $p$ then $q$. The alias calculus rule is:

$$a \text{ » } (p \,;\, q) \quad = (a \text{ » } p) \text{ » } q \qquad \textbf{/13/}$$

If the other rules of the calculus guarantee that $a \text{ » } p$ is an alias relation whenever $a$ is, this one also recursively yields an alias relation on the right side.

*Imprecision*: this rule introduces no imprecision.

*Alias diagram*: to carry out $p \,;\, q$, apply the transformations associated with $p$, then apply to the resulting graph those associated with $q$.

## 4.3 Forget

If $x$ is a variable, the notation **forget** $x$ denotes an instruction that removes any association of $x$ with any object. Corresponding programming language notations are:

```
x := Void          -- Eiffel
x = null;          -- C, Java etc.
```

(The reason for the special E0 syntax **forget** $x$ is that experience has shown that using assignment syntax, such that $x$ := **Void**, causes confusion with the regular form of assignment seen in 4.6 below, subject to a more general rule in the calculus.)

The rule is:

$$a \text{ » } (\mathbf{forget}\ x) \quad = \quad a \setminus\!\!- \{x\} \qquad \textbf{/14/}$$

with the operator $\setminus\!\!-$ as defined earlier (3.1).

*Imprecision*: this rule introduces no imprecision.

*Alias diagram*: to carry out **forget** $x$ on a diagram, remove $x$ from all edge labels that included it; to maintain the canonical form, also remove any edge that as a result goes down to a one-element label, as in this example:

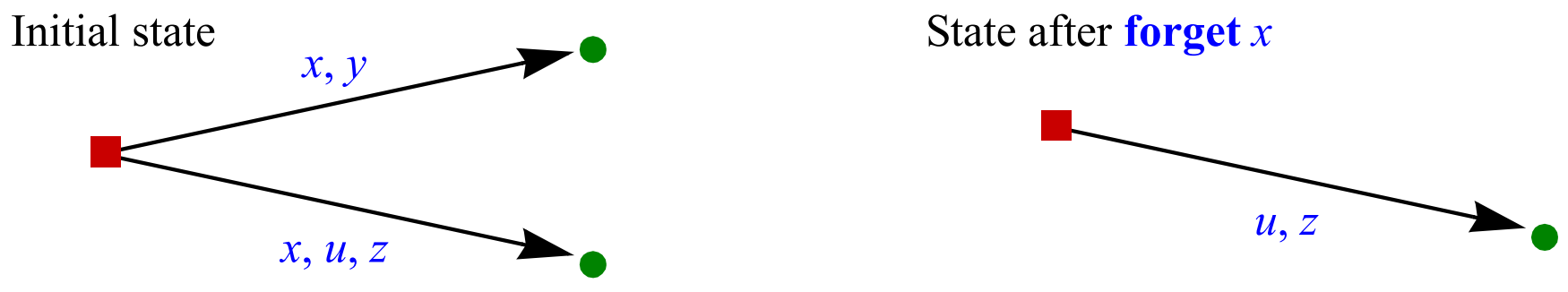

### 4.4 Creation

If $x$ is a variable, the notation **create** $x$ denotes an instruction that allocates a new object at a previously unused address. Corresponding programming language notations are:

| | |
|---|---|
| **create** *x* | -- Eiffel |
| x = new Type_of_x (); | -- C, Java etc. |

The effect on an alias diagram is the same as for **forget** $x$, and so is the rule:

$$a \gg (\textbf{create}\ x) \quad = \quad a \setminus\!\!- \{x\} \qquad \textbf{/15/}$$

*Imprecision*: this rule introduces no imprecision.

The **forget** and **create** instructions have different semantics — one removes all associations of a given variable with any objects, the other associates it with a new object — but in the alias calculus they are governed by identical rules.

### 4.5 Cut

If $x$ and $y$ are variables, the notation **cut** $x, y$ denotes an instruction that removes any aliasing between $x$ and $y$. It does not correspond to any common instruction of programming languages but, as noted in 2.2, will serve as an essential escape mechanism to remove undesired cases of imprecision in the calculus. The constructs

| | |
|---|---|
| **check** *x* /= *y* **end** | -- Eiffel |
| assert x != y; | -- JML, Spec# |

will be translated into **cut** $x, y$. (The semantics of **check** $p$ **end** in Eiffel is that the program is only valid with a proof that $p$ will always hold at the given program point; it is also possible for compilers that cannot perform such proofs to generate a run-time check that will stop the program if $p$ does not hold. The rules for assert in Spec# and JML are similar.)

The alias calculus rule is (with — denoting, as usual, set difference):

$$a \gg (\textbf{cut}\ x, y) \quad = \quad a - \overline{x, y} \qquad \textbf{/16/}$$

*Imprecision*: this rule introduces no imprecision.

*Alias diagram*: to carry out **cut** $x, y$, remove any edge with label $x, y$; replace any edge whose label includes $x, y$ and a non-empty set $A$ of other expressions by two edges, labeled $x, A$ and $y, A$, and leading to two separate nodes as shown in the next figure.

The need, in the second case, to replace an edge (and node) by two reflects the suggested practical use of **cut**: the operator lets us take advantage of finer-grain information, possibly coming from other sources, to improve the precision of the information provided by alias analysis. In the second case of the figure, the initial state conflated all of $x, u, z$ into a single alias class. Discovering that $x$ and $u$ are not related after all, we express this property by adding an instruction **cut** $x, u$, which separates the variables into two groups $\overline{x, z}$ and $\overline{u, z}$ listing $z$'s aliasing associations separately. The formal rule /16/ covers this semantics succinctly; it does not need to distinguish between the two cases illustrated by the diagram.

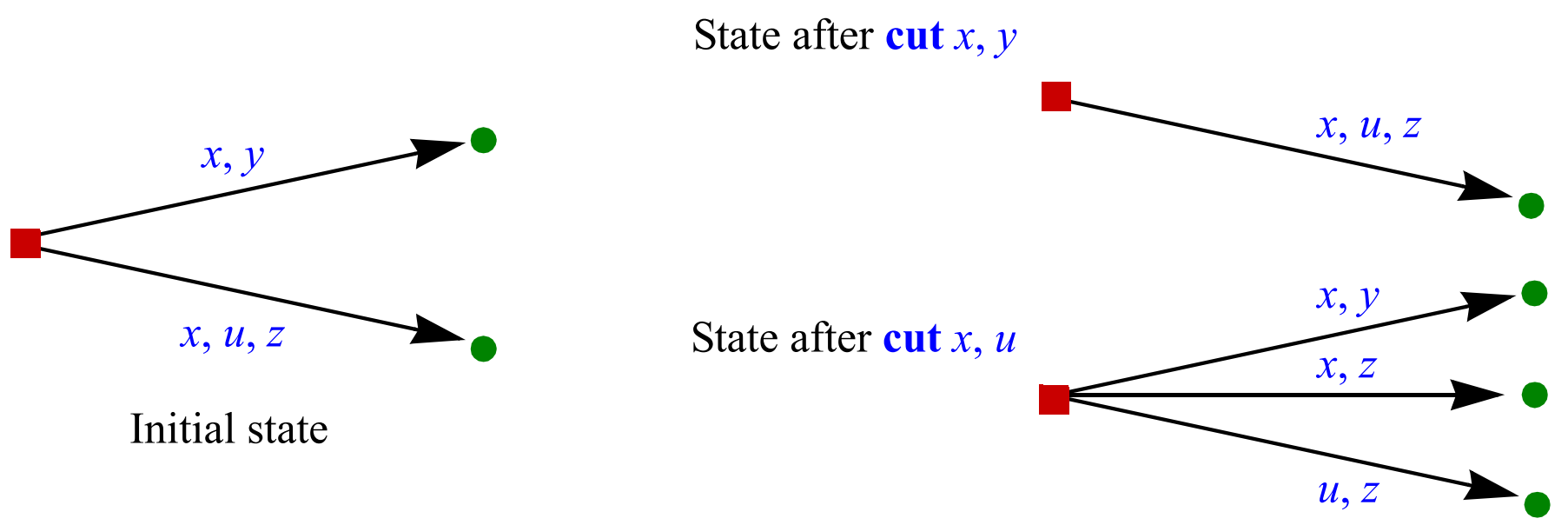

## 4.6 Assignment

The basic operation that creates alias pairs is assignment, written $x := y$. The rule is:

$$a \gg (x := y) \quad = \quad a\,[x\!: y] \qquad \textbf{/17/}$$

relying on the following operator involving an alias relation and two variables:

$$a\,[x\!: y] \quad = \quad \textbf{given } b \triangleq a \setminus\!\!- \{x\} \textbf{ then } \overline{b \cup (\{x\} \times (b \,/\, y))} \textbf{ end} \qquad \textbf{/18/}$$

The intuition behind this operator is that the assignment causes:

- Removal of any previous aliasing of $x$.
- Then, aliasing of $x$ to $y$ and to any other expression previously aliased to $y$.

The rule expresses this property. The relation $b$ is $a$ deprived of any pair involving $x$. The right side yields all the aliases not involving $x$, then adds the pairs $[x, u]$ where $u$ is in $b \,/\, y$, that is to say (/6/) either is $y$ or was aliased to $y$ in $b$, and applies the overline operator to symmetrize the relation and remove reflexive pairs (ensuring that the rule correctly handles the trivial case $x := x$).

*Example 1*: the value of $a \gg (z := f)$, where $a$ is

$$\overline{b, c, x},\ \overline{f, g, x},\ \overline{y, z} \qquad \textbf{/19/}$$

is (this example and all the following ones are as computed by the prototype implementation at se.ethz.ch/~meyer/down/alias.zip, on which the reader may try them):

$$\overline{b, c, x},\ \overline{f, g, x, z}$$

where $z$ has been removed from its previous association with $y$, then added to the associations of $f$.

*Alias diagram*: to carry out an assignment $x := y$ on an alias diagram, remove $x$ from all edge labels (removing the edge if the label goes down to zero or one element); if $y$ does not appear in any edge label, add a value node and an edge to it, labeled $y$; then add $x$ to any edge label containing $y$. The following figure shows some examples:

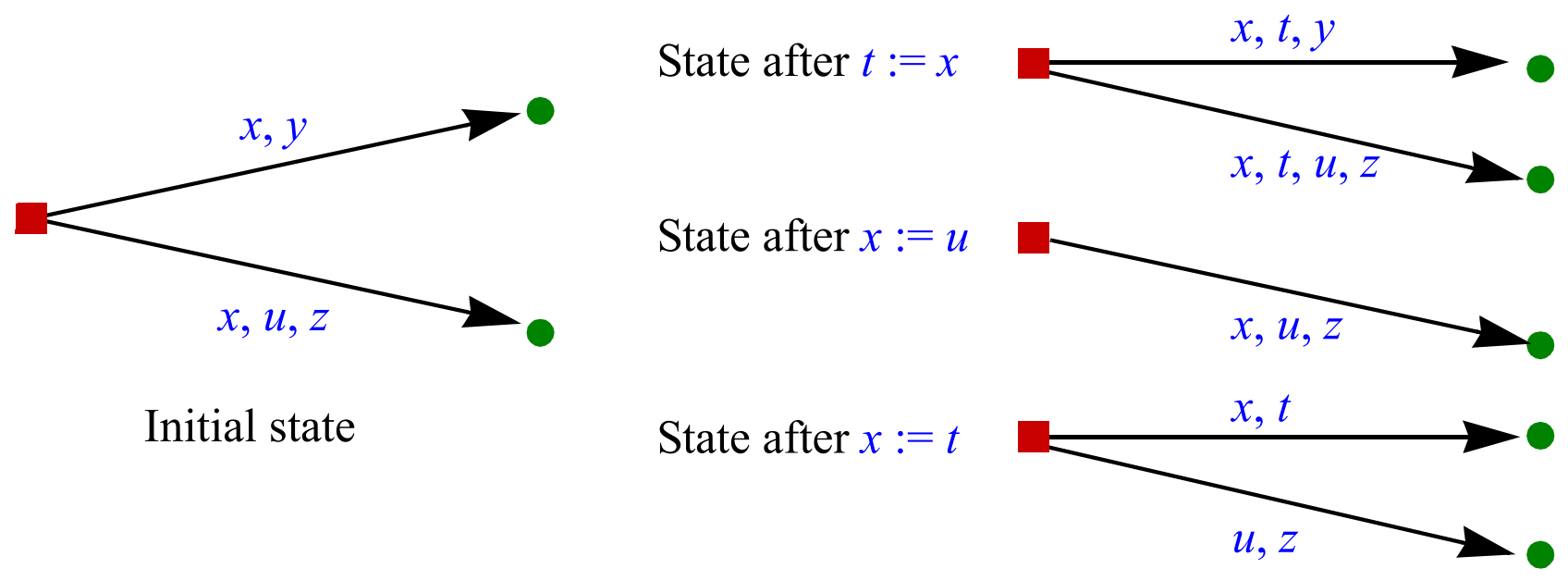

To deal with procedure calls, it is useful to generalize the $a\ [x\text{:}\ y]$ notation to the case in which $x$ and $y$ are not single variables but lists of variables:

$$a\ [x\text{:}\ y] \quad \stackrel{\Delta}{=} \quad (\ldots((a\ [x_1\text{:}\ y_1])\ [x_2\text{:}\ y_2])\ \ldots)\ [x_n\text{:}\ y_n] \qquad /20/$$

In other words, the list variant of the "**:**" operator applies the basic operator to the successive element pairs from both lists, assumed to be of the same size.

## 4.7 Conditional

E0 has a conditional instruction of the form

**then** $p$ **else** $q$ **end**

where $p$ and $q$ are instructions. The informal semantics of this instruction is that it executes either $p$ or $q$.

The rule is:

$$a \gg \textbf{then}\ p\ \textbf{else}\ q\ \textbf{end} \qquad = (a \gg p) \cup (a \gg q) \qquad /21/$$

The $\cup$ operator is here applied to two relations viewed as sets of pairs. As noted, $r1 \cup r2$ is an alias relation if both $r1$ and $r2$ are.

*Imprecision*: the conditional rule does not by itself introduce any imprecision, if we take the semantics of **then** $p$ **else** $q$ **end** to be that an execution can carry out either $p$ or $q$. In the translation of an ordinary programming language to E0, the source instruction would be **if** *cond* **then** $p$ **else** $q$ **end** for some condition *cond*. The condition is lost in translation; this may cause imprecision as in the earlier example (/3/).

*Example 2*: the program

**then** $x := b$ **else** $x := f$ ; $z := y$ **end** /22/

yields, when applied to $\overline{b, c},\ \overline{f, g}$, the alias relation $a = \overline{b, c, x},\ \overline{f, g, x},\ \overline{y, z}$ used as starting relation for the assignment example /19/.

*Note on the example*: the reader may wonder whether the assignment $z := y$ makes any sense without a prior assignment of a meaningful value to $y$. Such cases already arose in previous examples. For the alias calculus, however, this question need not alarm us, as it is a matter of convention for the underlying programming language. Some languages, such as the current void-safe version of Eiffel, guarantee that in any valid program $y$ will automatically be initialized on first use to a legal address, denoting an object. Alternatively, we may take the convention that every example program in this article implicitly starts with a sequence of **create** $x$ instructions, one for every variable $x$ appearing in the program. Or we could pass on the requirement to the programmer by including a static rule that disallows access before creation, in which case /22/ is invalid.

*Alias diagram*: to carry out **then** $p$ **else** $q$ **end**, produce two diagrams by separately applying $p$ and $q$ to the original diagram. Then combine the diagrams by retaining all their value nodes and all their edges. The result correctly represents the effect of the conditional but may not be in canonical form; make it canonical following the procedure seen in 3.2.

## 4.8 Repetition

E0 has an instruction

$$p^n$$

where $n$ is a natural integer. The semantics is that of **Skip** if $n = 0$ and otherwise, recursively, of $p^{n-1}$ ; $p$. Informally, this means $n$ executions of $p$.

The instruction is not important by itself (as only a few programming languages offer it directly) but as a stepping stone to the next construct, the loop instruction.

The rule is:

| | | | |
|---|---|---|---|
| $a \gg p^0$ | $= a$ | | **/23/** |
| $a \gg p^n$ | $= (a \gg p^{n-1}) \gg p$ | -- For $n > 0$ | **/24/** |

and is a direct consequence of the compound rule /13/.

*Imprecision*: the rule does not introduce any imprecision.

*Example 3 to 8*: take $x := y \ ; y := z \ ; z := x$ for $p$ and $\overline{c, y},\ \overline{d, z}$ for $a$. Then:

| | | | |
|---|---|---|---|
| $a \gg p^0$ | $=$ | $\overline{c, y},\ \overline{d, z}$ | $= a$ |
| $a \gg p^1$ | $=$ | $\overline{c, x, z},\ \overline{d, y}$ | |
| $a \gg p^2$ | $=$ | $\overline{c, y},\ \overline{d, x, z}$ | $= a$ |
| $a \gg p^3$ | $=$ | $\overline{c, x, z},\ \overline{d, y}$ | $= a \gg p^1$ |
| $a \gg p^4$ | $=$ | $\overline{c, y},\ \overline{d, x, z}$ | $= a$ |
| etc. | | | |

The sequence oscillates indefinitely, for odd and even $n$, between the values of $a \gg p^0$ (which is $a$) and $a \gg p^1$. This is as intuitively expected since $p$ swaps the values of $y$ and $z$.

## 4.9 Loop

The E0 instruction

| |
|---|
| **loop** $p$ **end** |

has the informal semantics of executing $p$ repeatedly any number of times, including zero. Formally, if an instruction is defined as a relation between input and input states (see A.1), then **loop** $p$ **end** is simply $\bigcup_{n:\,\mathbb{N}} p^n$ .

A first form of the loop rule follows from this definition:

$$a \text{ » } \textbf{loop}\ p\ \textbf{end} = \bigcup_{n:\,\mathbb{N}} (a \text{ » } p^n) \qquad \textbf{/25/}$$

*Imprecision*: the rule by itself does not introduce any imprecision. Imprecision may follow, however, from translating loop constructs as found in actual programming languages into the E0 form, since the translation will lose any information that the programmer or prover may have about the actual number of iterations, deduced for example from the loop exit condition in the usual **while** or **until** form of loop.

*Theorem*: the alias relation induced by a loop per /25/ is finite.

*Proof*: trivial since alias relations are members of $\mathbb{P}(E \times E)$ for a finite set $E$ (of variables and expressions appearing in a program); even an infinite union of such relations is finite. ■

This theorem, and the loop rule in its first form /25/, are not directly useful since they do not yield a practical way of computing $a$ » **loop** $p$ **end**. A more interesting version of the theorem, the loop aliasing theorem, will follow from the discussion of monotonicity appearing next, and will yield the practical version of the loop rule, /29/ below.

## 4.10 Monotonicity and the loop aliasing theorem

To deal effectively with loops, and procedures as introduced next, we need structural properties. For any instruction $p$, we define monotonicity of the » operator, with respect to the partial order relation $\subseteq$ (here over relations, that is to say, subsets of $E \times E$), as the following property for any alias relations $a$ and $a'$:

$$a \subseteq a' \quad \Rightarrow \quad (a \text{ » } p) \subseteq (a' \text{ » } p) \qquad \textbf{/26/}$$

*Alias monotonicity theorem*: all rules given so far satisfy monotonicity.

*Proof*: the rules for the control structures — compound, conditional, repetition and loop — clearly preserve monotonicity if the constituent instructions satisfy it; so we must establish monotonicity for basic instructions. Since $a$ » $p$ is deduced from $a$, and $a'$ » $p$ similarly from $a'$, by some additions and removals of pairs, the proof must show that any pair added to $a$ is also added to $a'$ and that any pair removed from $a'$ either was not in $a$ or is also removed from $a$. The only direct source of additions is the assignment rule /17/; added pairs for the assignment $x := y$ include $[x, y]$, which will also be added to $a'$, and $[x, z]$ where a pair $[y, z]$ was in $a$, and hence in $a'$, so that this pair will be added to $a'$. Removal of pairs occurs through the rules for **forget**, **create**, **cut** and assignment. In the first three cases the set of removed pairs depends entirely on the instruction and not on $a$ or $a'$: removing any of the pairs from $a'$ will remove it from $a$ if it was there. In the assignment case, the rule removes all pairs $[u, v]$ where either $u$ or $v$ is $x$; if any such pair in $a'$ is also in $a$, it will be removed from $a$. The rule also removes all reflexive pairs, but none of those comes from the original $a$ or $a'$ as they are alias relations. ■

The following properties are also of interest:

$$((a \gg p) \cup (a' \gg p)) \;=\; (a \cup a') \gg p \qquad /27/$$
$$(a \cap a') \gg p \;=\; ((a \gg p) \cap (a' \gg p)) \qquad /28/$$

In each case the left side is a subset of the right side as a consequence of the alias monotonicity theorem. The proof of the reverse inclusions follows, as for that theorem, from considering additions and removals for each kind of instruction.

The next theorem yields a practical way to compute the alias relation induced by a loop:

**Loop aliasing theorem**

For given $p$, let the sequence $t$ be defined by $t_0 = a$ and $t_{n+1} = t_n \cup (t_n \gg p)$. There exists an integer $N$ such that

1 For any $i < N$, $t_i \neq t_{i+1}$.

2 For any $i > N$, $t_i = t_N$.

3 $t_N = (a \gg$ **loop** $p$ **end**$)$.

The proof of this theorem is longer than one might expect and appears in appendix C. As a consequence of the theorem we will use the following version of the loop rule:

$a \gg$ **loop** $p$ **end** $\quad = \quad t_N$ /29/

-- For the first $N$ such that $t_N = t_{N+1}$,

-- with $t_0 = a$ and $t_{n+1} = t_n \cup (t_n \gg p)$.

*Example 9*: a loop with the same body as in the repetition example

```
loop x := y ; y := z ; z := x end
```

and started with the same initial alias relation $a = \overline{c, y},\ \overline{d, z}$ reaches its fixpoint at $t_2$:

$t_0 = a = \overline{c, y},\ \overline{d, z}$

$t_1 = \overline{c, x, z},\ \overline{c, y},\ \overline{d, y},\ \overline{d, z}$

$t_2 = \overline{c, x, z},\ \overline{c, y},\ \overline{d, y},\ \overline{d, x, z}$

$t_3 = t_2$

-- etc. (all subsequent values equal to $t_2$).

In this example, the sequence $a \gg p^n$ did not converge, as we saw in 4.8. But the loop aliasing theorem tells us that the sequence $t_n$ always reaches a fixpoint — here $t_2$ — finitely.

### 4.11 A more intricate example

*Example 10*: as a more extensive application of the E0 calculus, involving instructions of all the kinds encountered so far, consider the following program $p$ (semicolons omitted at end of lines):

```
then x := y else x := a end
then cut x, y ; z := x else end
g := h ; x := y ; z := a; b := x
loop e := f ; a := e end
loop
      then c := b ; a := f ; g := x else c := a ; a := g end
      f := x
end
b := z ; forget b ; a := e ; create z ; a := h ; cut a, g ; create x
```

The calculus yields, as the value of $\varnothing$ » $p$, the relation $\overline{a, c, h}$, $\overline{c, e, f}$, $\overline{c, f, g, y}$, $\overline{c, g, h}$.

### 4.12 Formalizing E0 and soundness

It is natural at this point to ask how the calculus can be justified in terms of a definition of the programming language. Appendix A (section A.1) gives a proof of soundness, for the main instructions of E0, with respect to a formal definition of the language.

## 5 Introducing procedures

Our next language, E1, adds to E0 the notion of procedure, with possible arguments. A procedure $r$ is defined by:

- A procedure name.
- A procedure body, written $\underline{r}$.
- A list of formal arguments, written $r^\bullet$.

E1 has a new instruction, **call** $r$ ($l$), where $r$ is a procedure and $l$ a list of variables; the effect is to execute $\underline{r}$ with $l$ as actual arguments; for empty $l$, the parentheses are omitted. (In actual programming languages the usual syntax for such a call is just $rn$ ($l$) where $rn$ is the name of $r$.) A program is defined by a non-empty set of procedures and the specification of one of them, with no arguments, as the main procedure.

The rule for a program $pr$ with main procedure $Main$ is

| | | | |
|---|---|---|---|
| $a$ » $pr$ | = | ($a$ » **call** $Main$) | **/30/** |

used in practice with $\varnothing$ for $a$, assuming every program starts with an empty alias relation. The rule for the call instruction itself is:

| | | | |
|---|---|---|---|
| $a$ » **call** $r$ ($l$) | = | $a\,[r^\bullet : l]$ » $\underline{r}$ | **/31/** |

using the ":" operator (generalized to lists of variables /20/ from the original definition for assignment /18/). This rule indicates that we understand the effect of a procedure call as the execution of the procedure's body, preceded by a sequence of assignments $f_i := u_i$ for every formal argument $f_i$ and the corresponding actual argument $u_i$.

In the absence of mutually recursive procedures, computing the alias relation of a program can simply proceed as in the previous examples: for every program element $p$, starting with the entire program, apply the corresponding alias calculus rule, which expresses $a \text{ » } p$ in terms of $a' \text{ » } p'$ for sub-elements $p'$ of $p$; the process terminates when applied to atomic elements such as assignments. This scheme no longer directly works for a program that includes mutually recursive procedures, since the computation of $a \text{ » } \underline{r}$ through the call rule /31/ may lead to a new evaluation of $a$ » **call** $r\ (l')$. To obtain a directly applicable process, we note that if a program consists of a number of procedures $r_1$, $r_2$, … $r_n$, and use the notation $b_i\ (a)$ for $a \text{ » } \underline{r}_i$ ,we may write the application of the call rule to any one of them, expanding $a \text{ » } \underline{r}_i$, as

$$b_i\ (a) \quad = \quad AL_i\ (b_1\ (f_{1,\, i}\ (a)),\ b_2\ (f_{2,\, i}\ (a)),\ \ldots\ b_n\ (f_{n,\, i}\ (a)))$$

where all the functions involved, $AL_i$ and $f_{j,\, i}$, deduced from applying the rules of the calculus to the text of $b_i$, are monotone. If $r_1$ is the main procedure, defining the alias relation induced by the whole program, computing $b_1\ (\varnothing)$ will give us, in the resulting $b$ vector, the alias relation at the exit point of every procedure (which is where we need it to apply axiomatic semantics, for example in weakest-precondition style). Since all functions involved are monotone and the set of relations is finite, standard reasoning shows that starting with empty relations for all the $b_i$ and iterating will reach a fixpoint finitely, yielding the desired result. The prototype implementation directly applies these ideas, as illustrated by the next example.

*Imprecision*: by itself this rule introduces no imprecision. Translations from programming languages will cause imprecision because E1 treats every formal argument as a global variable; local variables and function results will also be treated as global. The translation will lump together, for the computation of aliases of a local variable, result or formal argument, values that belong to different recursive incarnations of a given recursive routine (or to concurrent executions of that routine in different threads)

*Example 11* (in this example and the following ones the starting alias relation is empty): we consider the recursive procedure

```
procedure Main
    then
        x := y
    else
        x := a ; call Main
    end
end
```

The resulting relation is just $\overline{x, y}$: the conditional's second branch can never contribute anything.

*Example 12*: If we reverse the order of the instructions in the else clause of the previous example (giving **call** *Main* ; $x := a$), we get $\overline{a, x}$, $\overline{x, y}$.

*Example 13*: the following are mutually recursive procedures (without arguments and still simple, to allow intuitive manual verification of the result):

```
procedure Main
    then x := y else x := a ; call q end
end
procedure q
    x := b ; then call Main else a := c end
end
```

The result, with *Main* as the main procedure, is $\overline{a, c}$, $\overline{b, x}$, $\overline{x, y}$. In particular, $x$ can get aliased to $a$ and $a$ to $c$, but not $x$ to $c$.

*Example 14*: another case of mutually recursive procedures:

```
procedure Main
    then x := y else x := a end
    then cut x, y ; z := x else end
    then call q else g := h end
    x := y ; z := a ; b := x
    loop
        e := f
        then a := e else end
    end
    then c := b ; a := f ; g := x else
        loop c := a ; a := g end
        call Main
    end
    f := x ; b := z; forget b ; a := e ; create z; a := h
    cut a, g ; create x
end
procedure q
    then m := n else m := h ; call Main end
end
```

The result is $\overline{a, h, m}$, $\overline{c, e, f, g, y}$, $\overline{m, n}$. This example is not drawn from any actual program but illustrates the application of the calculus to procedures with a complex recursion and control structure.

## 6 The object-oriented calculus

The next and last language level, E2, introduces object-oriented mechanisms. E2 is sufficiently powerful to support applying the calculus to a modern object-oriented language such as Eiffel, Java or C#. The relevant part of object technology here is the dynamic object model: dynamic object creation, pointers or references (we will consider the two terms synonymous), and the possibility for objects to contain pointers to other objects. This last facility is the only novelty of E2's dynamic model, since E0 and E1 already offered the first two.

Other object-oriented mechanisms such as inheritance and genericity have only marginal influence on aliasing. Polymorphism and dynamic binding introduce some interesting issues which another article will address.

## 6.1 New language concepts

Making E2 support object-oriented programming means adding three language concepts:

- Qualified expressions with two or more components, such as *x.y.z*, which can be used as sources of assignments, as in *u* := *x.y.z*.
- Qualified calls, such as *x.f* (*u*, *v*).
- The notion of current object (**Current** in Eiffel, this in C++ and Java, self in Smalltalk). This is the central concept of object technology, giving rise to the "*general relativity*" principle of O-O programming: every operation is relative to a current object; starting a qualified call *x.f* (*v*) makes a new object (the object attached to *x*) current; ending such a call restores the previous current object as current.

We will not directly consider qualified assignments of the form *x.a* := *v* permitted by programming languages such as Java, C# and C++. It may be possible to include qualified assignments directly into the theory, a task that the present article does not undertake (as a matter of principle, since qualified assignments fly in the face of all principles of software engineering, and even the designers of languages that include this mechanism advise against using it); it happily leaves it for other authors to solve. The omission of this mechanism in the theory and calculus as described here has no practical consequence on the application to the relevant programming languages, since it suffices to assume a pre-processing step that translates all qualified assignments *x.a* := *v* into qualified calls to setting procedures, such as *x.set_a* (*v*).

## 6.2 Object-oriented alias diagrams

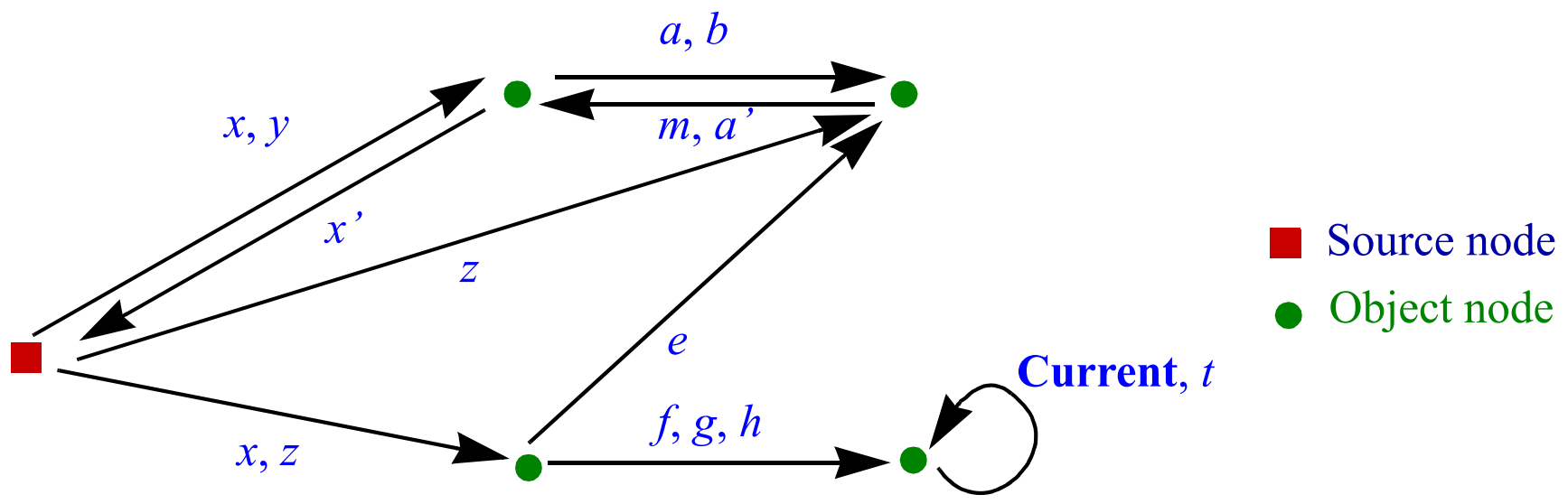

The new expressions forms appearing in the figure, **Current** and "negative references" such as *x'*, will be defined shortly. E2 alias diagrams still have a source node, which now represents the current object, but that node no longer has any special property; edges can exist between value nodes (from now on called object nodes), as illustrated by the figure. As this example suggests, cycles are now possible, albeit between objects nodes only. Cycles involving negative references will play a fundamental role in representing object-oriented concepts; we will see that they arise as a result of passing arguments to qualified calls.

One-expression edge labels 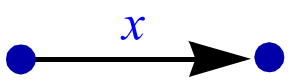

, which we discarded in E0 and E1, are meaningful for O-O alias diagrams. Also, we no longer systematically remove the end node when we remove an edge, but only do so if no other edge leads to that node. (This property reflects the need for garbage collection in an object-oriented model.)

An object node represents a set of possible objects, all of the same type (class); the interpretation of an edge with labels $x, y\ldots$ between two object nodes, representing sets of objects *OS1* and *OS2*, is that every object in *OS1* may have reference fields to an object in *OS2*; since in typed object-oriented programming every field of an object corresponds to an attribute (also called "member variable" or "data member" in various O-O languages) in the relevant class, the fields involved are those corresponding to attribute names.

In the figure, $x$, $y$ and $z$ are names of attributes of the current class; $e$, $f$, $g$ and $h$ are attributes of the class describing the object in the middle-bottom node. The calculus does not need information about the classes; we assume that it is applied to a typed O-O language after type checking, so that every attribute name refers unambiguously to a class. This convention is particularly important in Eiffel where style rules suggest the systematic use, for consistency, of a small set of feature names such as *first and* item. In the application of the calculus to a specific programming language, a good convention might be to identify the class as part of the attribute name, as in $item_{LIST}$, $item_{CELL}$ etc. We will need no such convention here. The nodes in the last figure might correspond to objects of the same type or different types.

In previous language levels the set $E$ of expressions only contained single variables. E2 offers three new forms of expression:

- The special expression **Current** represents the current object (relative to any node). Informally, **Current** denotes a link from a node to itself, as in the bottom-right node of the last figure.
- For any variable $x$, the **negation** (or "inverse") of $x$ is written $x'$. Informally, consider a call $x.r$, executed on behalf of a certain client object, which applies $r$ to a supplier object referenced by $x$; then $x'$ represents a reference back from the supplier to the client. It will appear in edges between the corresponding nodes, as in the preceding figures. Like **Current**, the negation operator introduces cycles into E2 graphs.
- Finally, E2 supports dot expressions of the form $x.y.z\ldots$

The presence of dot expressions gives alias diagrams a richer meaning: aliases arise not only from edges but also from *paths* in the diagram. The rule is that if two paths have the same starting node and the same ending node, the corresponding dot expressions are aliased. Consider for example, in the last diagram, the edge labeled $z$ from the source node to the top-right node; it implies that $z$ is aliased to $x.a$, $x.b$, $y.a$ and $y.b$ (paths through top nodes) as well as $x.e$ and $z.e$ (bottom paths).

Section A.2 (part of an appendix) shows how to extend the formal model defined for previous language levels to the object-oriented mechanisms just introduced.

### 6.3 Dot completeness

For simplicity it is convenient to add the dot to the calculus as an operator on variables and expressions representing paths: if $v$ is a variable and $e$ an expression $x.y.z\ldots$, we write $v.e$ to denote the path $v.x.y.z\ldots$ and extend this notation to two expressions, writing $e.f$ for the concatenation of $e$ and $f$.

The following fundamental property, reflecting the preceding observation on alias diagrams, characterizes the semantics of aliasing with dot expressions:

**Dot completeness**

An alias relation $a$ involving dot expressions must satisfy, for any expression $e1$, $e2$, $f1$ and $f2$:

$$[e1, e2] \in a \wedge [f1, f2] \in a \quad \Rightarrow \quad [e1.f1, e2.f2] \in a \qquad \textbf{/32/}$$

This requirement is added to the basic definition of alias relations as symmetric and irreflexive (3.1). All the calculus rules to be introduced now will preserve dot completeness. This is also trivially the case with the previous rules, which did not introduce dot expressions.

In the dot calculus, **Current** plays the role of zero element and variable negation the role of the negation operation. For any expression $e$ (a variable or a path) and any variable $x$:

| | | | |
|---|---|---|---|
| **Current**$.e$ | = | $e$ | **/33/** |
| $e.$**Current** | = | $e$ | **/34/** |
| $x.x'$ | = | **Current** | **/35/** |
| $x'.x$ | = | **Current** | **/36/** |

and as a consequence, for non-empty $e$:

| | | | |
|---|---|---|---|
| $x.x'.e$ | = | $e$ | **/37/** |
| $x'.x.e$ | = | $e$ | **/38/** |

/33/ and /34/ express that **Current** always represent a link to the current node. Note that the interpretation of **Current**, like everything else in the general relativity of object-oriented programming, pertains to an object and the corresponding class; /34/ describes a situation such as

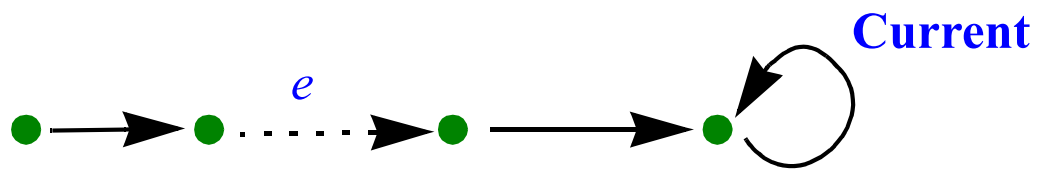

where the various nodes involved might correspond to different classes. In the application to a typed object-oriented language, **Current** is really $\textbf{Current}_C$ for some class $C$. Clearly, $e.\textbf{Current}_C$ only makes sense if $C$ is the class of the objects reached by $e$ (the rightmost node in the figure); the alias calculus need not concern itself with this question, since we assume it is applied to type-checked programs.

In this framework, the alias calculus needs only two more rules to account for object-oriented programming: an adaptation of the assignment rule to account for multidot sources; and a rule for qualified calls **call** $x.r$.

## 6.4 Dot expressions as sources of assignments

In an assignment $x := y$, the source expression $y$ may now be a multidot expression, such as $u.v.w$. An illustration with an example alias diagram (with no initial aliasing) is:

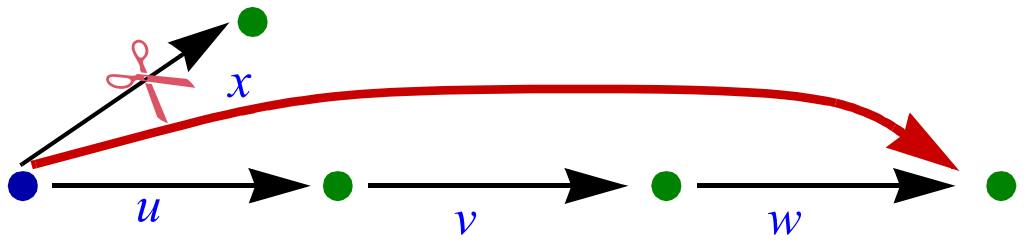

The original assignment rule /17/ only requires a small adaptation. In fact the rule itself, which reads $a \text{ » } (x := y) = a\ [x{:}\ y]$ does not change; nor does the definition of $a\ [x{:}\ y]$:

$$a\ [x{:}\ y] \quad = \quad \textbf{given } b \triangleq a \setminus\!\!- \{x\} \textbf{ then } \overline{b \cup (\{x\} \times (b \,/\, y))} \textbf{ end} \qquad \text{-- Same as /18/}$$

We need, however, to adapt the definition of the $\setminus\!\!-$ operator to account for possible dots in $y$. The original definition (3.1) was that $r \setminus\!\!- A$ is $r$ deprived of any pair that involves a member of $A$. From now on $r \setminus\!\!- A$ is also deprived of any pair involving a multidot expression whose first component (in the sense of $u$ in $u.v.w$) is a member of $A$.

As a consequence, the set $\overline{b \,/\, y}$ (involved in the set of pairs $\overline{\{x\} \times (b \,/\, y)}$, added to the relation on the second line above) may be empty, in which case $\overline{\{x\} \times (b \,/\, y)}$ is itself empty. This reflects an important practical property: while in the non-O-O calculus an assignment $x := y$ always adds the pair $[x, y]$ to the alias relation, this is not necessarily the case with dot expressions. In the assignment

```
x := x.a
```

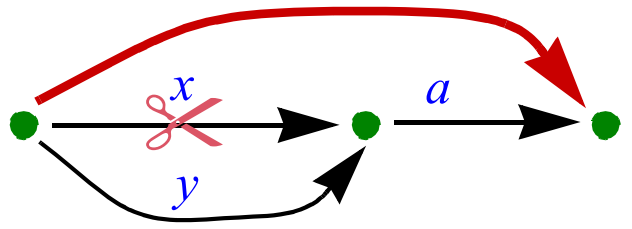

we should **not** alias $x$ to $x.a$! The instruction removes all aliases of $x$, and creates no new aliasing unless $x$ was previously aliased to some other expressions; then for every such expression $y$, it aliases $x$ to $y.a$.

These observations do not rule out the possibility for $x$ to become aliased to $x.a$; although such a case cannot be the result of the assignment above, it will happen if $a$ is aliased to **Current**.

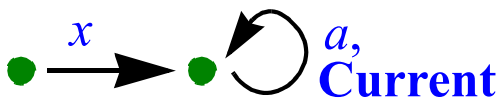

The rule captures all these cases.

*Imprecision*: the rule introduces no imprecision.

*Example 15*: the following program uses dot expressions as assignment sources:

```
x := y ; a := b
z := x.a ; x := x.a
```

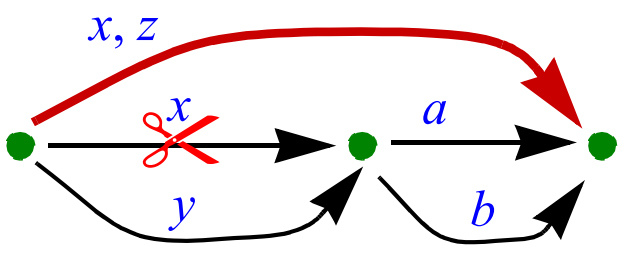

The result is $\overline{a, b}$, $\overline{x, y.a, z}$, $\overline{x, y.b, z}$.

### 6.5 Qualified call

The last remaining construct is the qualified procedure call **call** $x.r\ (l)$ where $l$ is a list of actual arguments. To handle it in the alias calculus, we need the following notation: if $a$ is a relation (in our examples, an alias relation), $x \bullet a$ denotes the relation containing all pairs $[x.e, x.f]$ such that $a$ contains $[e, f]$.

In a naïve approach to handling $x.r$, we would note that if a call to $r$ (unqualified) aliases $e$ to $f$ then a call to $x.r$ aliases $x.e$ to $x.f$. Then $a \gg x.r$ would be $x \bullet (a \gg r)$. This does not, however, capture the possible changes to aliasing on the side of the client (the object on whose behalf the call $x.r$ is made). Consider for example, in an object-oriented programming language, the instructions

$l := m$
$x.r\ (l)$

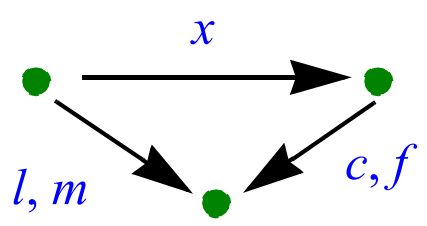

(*See below the completion of this figure*)

with, in the supplier class (the type of $x$):

$r\ (f : T)$ **do** $c := f$ **end** /39/

where $c$ is an attribute of that class. The execution of the call starts with the equivalent of an assignment $f := l$ of the actual argument to the formal argument. The naïve rule would give us the (symmetrized) pairs $[l, m]$, $[c, f]$ and $[x.c, x.f]$, which are correct, as well as $[f, l]$ and as a result $[f, m]$ which are meaningless since they involve variables applicable to different objects (and possibly different classes). It misses, on the other hand, the aliasing of $x.c$ to $l$ and $m$. It is unsound.

Obtaining a sound rule requires the **negation operator** on references. The correct way to represent actual-formal argument association for **call** $x.r\ (l)$ is not $f := l$ but

$f := x'.l$

which associates the formal argument $f$ to the actual argument $l$ *considered in the context of the supplier object*. The role of the negative reference $x'$ is to provide, in a qualified call, a link back to the client. This enables the supplier, if needed, to update references that belong to the client side — a principal facility, although one fraught with obvious risks (aliasing risks in particular), of the object-oriented style of programming.

The reason for the negation rules /35/ and /36/ should now be clear: $x.x'$ is **Current** (for the client) and $x'.x$ is **Current** (for the supplier).

The rule given below follows from this discussion, with some further adaptations:

- In the general case the actual argument $l$ is not a single expression but a list. It will be associated with the list of formal arguments, written $r^\bullet$ (section 5).
- The actual-formal association should cease after the call. In the last example, the alias pair $[l, x.f]$ is applicable throughout the execution of the call, but should disappear on completion of this call. This indicates that after applying the techniques described the rule must remove all pairs involving an element of $x \bullet r^\bullet$.

The alias calculus rule for qualified calls is:

$$a \gg \textbf{call}\ x.r\,(l) \quad = \quad x{\scriptstyle\blacksquare}\,((x'{\scriptstyle\blacksquare}\, a)\ [r^{\bullet}: x'{\scriptstyle\blacksquare}\, l]) \gg \underline{r}) \backslash\!\!- x{\scriptstyle\blacksquare}\, r^{\bullet} \qquad \textbf{/40/}$$

The final term $\backslash\!\!- x{\scriptstyle\blacksquare}\, r^{\bullet}$ represents the removal of pairs involving $x.f$ for a formal argument $f$, as just discussed. The expression $(x'{\scriptstyle\blacksquare}\, a)\ [r^{\bullet}: x'{\scriptstyle\blacksquare}\, l]$ represents the result, starting from the original relation $a$, of prefixing both elements of every pair by $x'$ and a dot, then adding pairs $[f, x'.c]$ for every formal argument $f$ and the corresponding actual $c$.

For a routine without arguments, these two elements disappear, giving:

$$a \gg \textbf{call}\ x.r \quad = \quad x{\scriptstyle\blacksquare}\,((x'{\scriptstyle\blacksquare}\, a) \gg \underline{r}) \qquad \textbf{/41/}$$

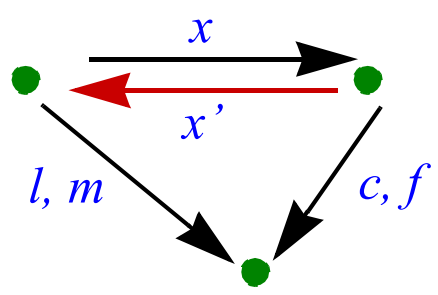

(*Completion of preceding figure*)

We can use this simplified form to see how the general rule works. To compute the aliasings induced by **call** $x.r$ starting from the aliasing environment $a$, we need to compute the aliasings induced by an execution of the procedure body $\underline{r}$; not exactly from $a$, however, since $a$ is relative to the client object, but from $a$ as seen by the supplier object (known through $x$, the target of the call). This supplier view is $x'{\scriptstyle\blacksquare}\, a$, with both elements of every pair in $a$ prefixed by the negative reference $x'$, a back pointer giving access to the client. In the above example /39/, $r$ executes $c := f$ using a formal argument $f$ aliased to the actual argument $l$, known in the routine as $x'.l$ (through the actual-formal argument reflected by the expression $a\ [r^{\bullet}: x'{\scriptstyle\blacksquare}\, l]$ in /40/). Then $c$ will get aliased to $x'.l$. The resulting supplier-side relation has among others the pairs $[f, x'.l]$ and $[c, x'.l]$.

This relation, which we may call $a'$, is $(x'{\scriptstyle\blacksquare}\, a) \gg \underline{r}$. It is only meaningful in the environment of the supplier. After the execution of $r$ returns, we need to interpret the resulting aliases in the environment of the client. Since the client knows the supplier as $x$, the relation we need is $x{\scriptstyle\blacksquare}\, a'$, that is to say $a'$ but with both elements of every pair prefixed by $x$. If such an element is of the form $x'.e$, this prefixing will yield just $e$, since the dual rule **/35/** tells us that $x.x' =$ **Current** and **/33/** tells us that **Current**$.e = e$. In the example, the pair $[c, x'.l]$ in $a'$ will give $[x.c, l]$; this is the proper result as illustrated. Note that the same process applied to the pair $[f, x'.l]$ will give $[x.f, l]$; such an association with a formal argument of $r$ is no longer meaningful after the routine's execution, and is removed by the last part of the full rule /40/.

Thus we are permitted to prove that the unqualified call creates certain aliasings, on the assumption that it starts in its own alias environment but has access to the caller's environment through the negated variable, and then to assert categorically that the qualified call has the same aliasings transposed back to the original environment. This change of environment to prove the unqualified property, followed by a change back to the original environment to prove the qualified property, explains well the aura of magic which attends a programmer's first introduction to object-oriented programming[3].

---

3. Since no reader of the original version of this article seemed to notice the reference, it may be appropriate to reveal that this paragraph is a re-rendering of Tony Hoare's comments on the axiomatic model of recursion in his 1971 *Procedures and Parameters: An Axiomatic Approach*.

In the example, as illustrated by the figure, the resulting alias relation (for the original object, represented by the top-left node in the last figure) is $\overline{l, m, x.c}$. It mentions neither the negated variable *x'* nor the formal argument *f*, which are meaningful only for the target object (top-right in the last figure).

To update the previous proofs that a fixpoint exists and is reached finitely, we note that applying **call** *x*.*r* to an alias relation *a* may increase the dot count of at least one of element of some pairs in *a*. In the case of recursive or mutually recursive procedures, this property potentially invalidates the earlier finiteness arguments since the count may grow unbounded. It causes no practical problem, however, since the basic assumption of the theory of aliasing (2.3) is that it only considers expressions that actually appear in a program. So it suffices to limit application of rule /40/ to alias relations *a* whose dot count is no greater than the maximum dot count for expressions in the program, defining the dot count of a pair of expressions as the maximum of the dot counts of its elements, and the dot count of a relation as the minimum dot count of its pairs. (The precise argument is more subtle, since in principle two expressions of the program could become aliased as a result of rule /40/ aliasing each of them to an expression not appearing in the program and having a dot count higher than any that will be computed using the limited rule. It is easy to see, however, that this case is impossible.)

*Example 16*: the following program includes a qualified call with arguments, *x*.*q* **Current**, *f*).

```
            -- In root class:
procedure Main
      f := x.a            -- f is an attribute of the root class.
      call x.q (Current, f)
end

            -- In another class CC:
procedure q (b, c)
      d := b              -- d is an attribute of CC. (So is a.)
end
```

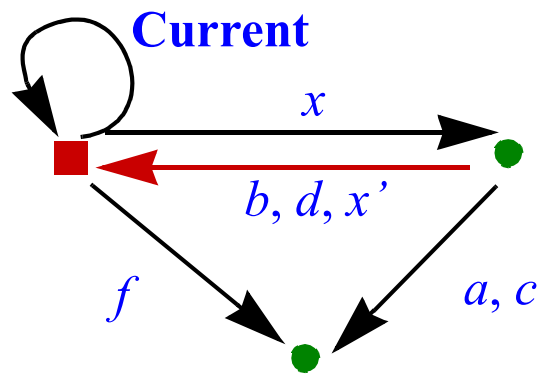

Applying the argument-passing scheme, as represented in /40/ by the term $a\,[r^{\bullet} : x' \centerdot l]$, is equivalent to pretending that the body of the routine *q* starts with assignments of actual to formal arguments:

```
b := x'.Current
c := x'.f
```

(*Example 17* is a variant of example 16 that handles argument passing explicitly, in this manner.)

The resulting alias relation is $\overline{Current, x.b, x.d}$, $\overline{f, x.a, x.c}$, $\overline{x.b.f, x.c}$. As appropriate, it only includes aliases reachable from the node representing the current object (the ■ node at the top-left in the figure). An alias pair such as [*a*, *c*], which applies to another node (the rightmost ● node in the figure, representing the target of the call *x*.*q*) appears as *x.a*, *x.c* in the alias relation relative to the current object node.

## 6.6 Aliasing among list structures

*Example 18*: for the final example (this variant and the next), consider the list manipulation program mentioned in the introduction. Here is the text again, including type declarations (for clarity only, since they do not affect the calculus) and the procedures from class *CELL*:

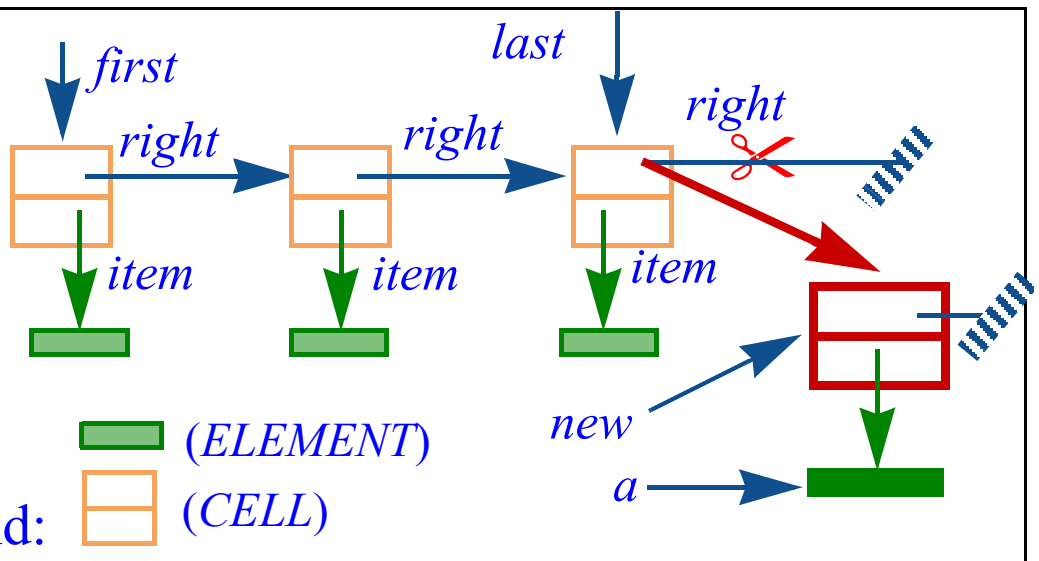

```
-- In class LIST:
   extend (a: ELEMENT)
            -- Add cell at end, with a as item.
      local
            new, last: CELL
      do
                  -- First create new cell:
            create new ; new.set (a)
                  -- Then insert new cell at end:
            if first = Void then
                        -- List was empty. It will now consist of a single cell, the new one:
                  first := new
            else
                        -- List was not empty; go to its last cell and make last denote it:
                  from last := first until last.right = Void loop last := last.right end
                        -- Then link last cell to new cell:
                  last.set_right (new)
            end
      end
-- In class CELL:
      set (v: ELEMENT)            do item := v end
      set_right (right: CELL)     do right := c end
```

Assume two separate lists *x* and *y*, to which we may add elements to our heart's content:

```
procedure build
      local
            el: ELEMENT
      do    -- The two instructions below could also be in separate branches of a then … else.
            extend_client := x ; loop create el ; call x.extend (el) end
            extend_client := y ; loop create el ; call y.extend (el) end
      end
```

Then we repeatedly access arbitrary elements of either list:

```
procedure Main
      call build
      f := x.first ; g := y.first
      loop then f := f.right else g := g.right end end
end
```

The alias relation (as obtained from running this example in the implementation) is:

$\overline{f, x.first, x.last}$, $\overline{f, x.first.right, x.last}$, $\overline{f, x.first.right.right, x.last}$, $\overline{f, x.last.right}$, $\overline{f, x.last.right.right}$, $\overline{g, y.first, y.last}$, $\overline{g, y.first.right, y.last}$, $\overline{g, y.first.right.right, y.last}$, $\overline{g, y.last.right}$, $\overline{g, y.last.right.right}$, $\overline{x.a, x.new.item}$, $\overline{x.last.right, x.new}$, $\overline{x.a, x.new.item}$, $\overline{y.a, y.new.item}$, $\overline{y.last.right, y.new}$

The full relation, as noted, would be infinite; it includes for example all pairs of the form [*x.first.right…., x.last*] with an arbitrary number of "*.right*" after *x.first*. As discussed in 6.5, the application of the theory to a particular annotated program breaks off at the highest dot length of expressions found in the program. To run the examples, the current implementation sets this maximum to thee dots, as illustrated in the above result.

The most important property of that result is that the relation does **not** include the pair [*f, g*], showing that no pointer in either list can ever become attached to a cell of the other:

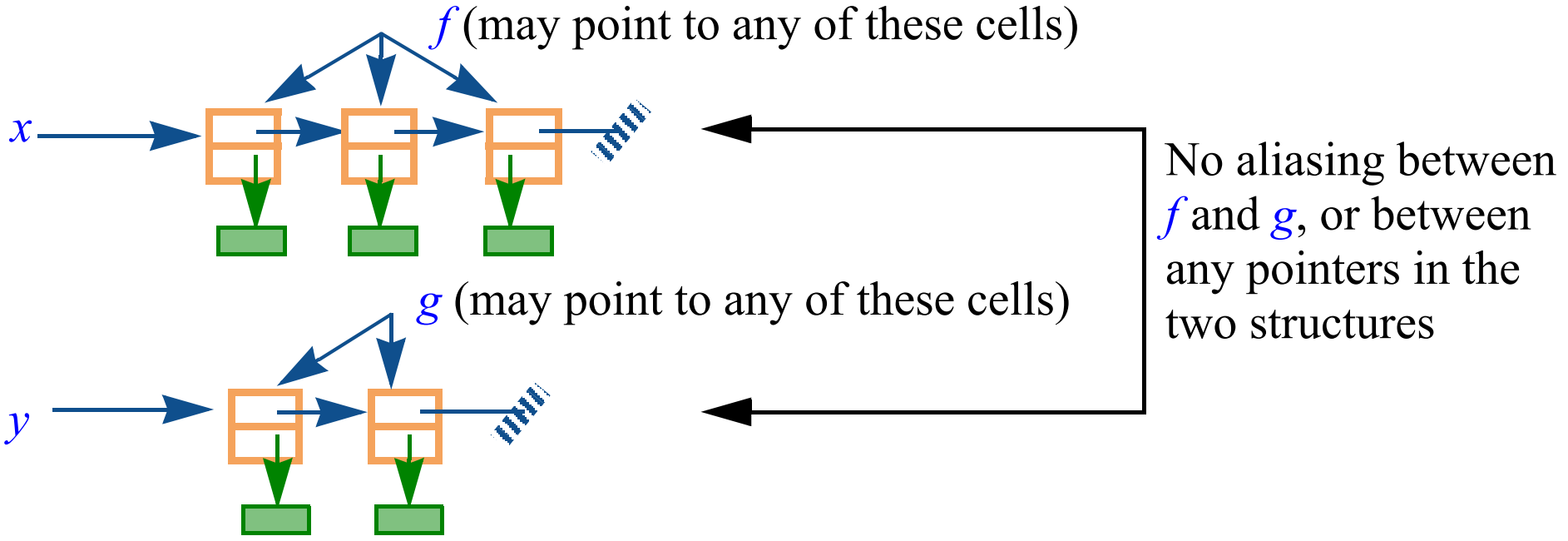

*Example 19*: The last example keeps example 18 unchanged with one exception: the assignment *x* := *y* added at the beginning of *Main*. The resulting alias relation now includes $\overline{f, g, x.first, y.first}$, $\overline{f, g, x.first.right}$ etc. (run the implementation to see the full list). The important property is that now, as a result of this single change, *f* **can** be aliased to *g*.

## 7 Prototype implementation

The prototype implementation is stand-alone, rather than integrated into the compiler of a programming language. It is written in Eiffel; the language's mechanisms of inheritance (particularly multiple inheritance), genericity and contracts have proved essential to the prompt completion of this implementation.

While a functional language might seem appropriate for such a prototype, the use of an imperative language was in fact essential. In particular, many delicate decisions involved when to duplicate a data structure, such as the representation of an alias relation, and when simply to update it. To complement standard object-oriented mechanisms, Eiffel's *agent* mechanism, which provides the power of closures in functional languages, also played a fundamental role.

The implementation makes it possible to write an E2 program and produce its alias relation in canonical form, as illustrated by the examples of this article. All the examples are part of the implementation and can be tried in the downloadable version.

At execution time response for these examples is immediate, but no complexity analysis has been performed to explore scalability.

# 8 Application to a programming language and open problems

The translation from an actual programming language involves the steps discussed earlier: ignoring conditions of conditionals and loops; replacing functions by procedures; replacing arguments, local variables and function results by attributes; associating negative references with targets of qualified calls.

A number of problems remain to be addressed:

- Although the existing implementation provides a convincing proof of concept, it should be integrated in the compiler for an actual programming language.
- Aliasing properties of arrays and other core data structures may require specific rules.
- The modular application of the calculus calls for special attention.
- Provision should be made for polymorphism and dynamic binding.
- On the theoretical side, the formal model and soundness proofs sketched in appendix A should be completed.
- A "must alias" variant of the calculus (appendix D) may be worth investigating.
- The application to the frame problem must be clarified (in a companion article).
- Application to large programs requires both experimentation and theoretical analysis of the algorithms' complexity.

## Appendix A: Formal model and soundness

This appendix shows how the calculus can be proved in reference to a formal definition of the programming language. It starts with the basic language (A.1) then sketches how to add object-oriented constructs (A.2). It also defines (B) $p\leftarrow$ (the set of variables that an arbitrary instruction $p$ may set), as needed by the definition of soundness /11/.

### A.1 Semantics and soundness of E0

An E0 program may be defined as a relation in $State \leftrightarrow State$. A deterministic language would use functions, possibly partial, rather than relations; non-determinism keeps the language definition simple, in particular for the loop construct.

A state $s$ is characterized by:

- A set of variables that have a value in that state: $s.def$ (a member of $\mathbb{P}(Variable)$).
- A set of addresses allocated in that state: $s.addr$ (a member of $\mathbb{P}(Address)$, assuming a suitable set $Address$).
- The values of the variables in the state, as represented by a function $s.value$, a member of $Variable \nrightarrow Address$ (using $\nrightarrow$ for the set of possibly partial functions), where **domain** $(s.value) = \{v\text{: } Variable \mid v \in s.def\}$.

To define a state $s$, it suffices to give $s.def$, $s.addr$ and $s.value$.

To define E0 formally we specify each instruction as a relation in $State \leftrightarrow State$, by considering in each case an arbitrary state $\sigma$ and stating the properties of states $\sigma'$ that may result from applying $p$. For example, in the case of **skip** (the identity relation on $State$), $\sigma' = \sigma$.

For the instruction **forget** $x$, the definition is: $\sigma'.def = \sigma.def - \{x\}$; $\sigma'.addr = \sigma.addr$; $\sigma'.value\ (y) = \sigma.value$ for $y \neq x$.

For **create** $x$, for some $na$ in $Addresses$ such that $na \notin s.addr$: $\sigma'.def = \sigma.def \cup \{x\}$; $\sigma'.addr = \sigma.addr \cup \{na\}$; $\sigma'.value\ (y) = na$; $\sigma'.value\ (y) = \sigma.value\ (y)$ for $y \neq x$.

For $x := y$: if $y \notin s.addr$, as for **forget** $x$; otherwise: $\sigma'.def = \sigma.def \cup \{x\}$; $\sigma'.addr = \sigma.addr$; $\sigma'.value\ (x) = \sigma.value\ (y)$ ; $\sigma'.value\ (z) = \sigma.value\ (z)$ for $z \neq x$.

For the compound $p\ ;\ q$: what this notation means as a mathematical convention, taken to denote composition of relations in the order given (the same as $q \circ p$).

All the elementary constructs defined so far are functions (deterministic). Non-function relations (representing possible non-determinism) may arise with:

- Conditional: **then** $p$ **else** $q$ **end** is defined simply as another notation for $p \cup q$.
- Loop: **loop** $p$ **end** is defined as $\bigcup_{n:\,\mathbb{N}} p^n$. The term $p^n$ (corresponding to the E0 repetition construct) retains its definition from mathematics: $p^0 =$ **skip**, $p^{n+1} = (p^n\ ;p)$.

In this framework, every state induces an alias relation defined as

$$\textbf{aliases}\ (\sigma) \triangleq \{[x, y] \mid x \in \sigma.def \wedge y \in \sigma.def \wedge \sigma.value\ (x) = \sigma.value\ (y)\}$$

An earlier formula /11/ defined soundness in an axiomatic semantics style. For a language such as E0, where instructions and programs are defined directly as relations, we may use the following version of the soundness rule, for any instruction $p$:

$$[\sigma, \sigma'] \in p \Rightarrow \ \textbf{aliases}\ (\sigma') \subseteq (\textbf{aliases}\ (\sigma) \gg p) \qquad \textbf{/42/}$$

As an example of soundness proof, consider **forget** $x$. For a given $\sigma$, the above definition of the **forget** instruction tells us that there is only one $\sigma'$ and that a pair $[y, z]$ is in **aliases** $(\sigma')$ if and only if $y \neq x$, $z \neq x$ and $\sigma.value\ (y) = \sigma.value\ (z)$. The pair is also in **aliases** $(\sigma) \gg$ **forget** $x$ since the **forget** rule /14/ defines **aliases** $(\sigma) \gg$ **forget** $x$ as **aliases** $(\sigma)$ \– $\{x\}$}.

In this example the $\subseteq$ relationship of the soundness requirement /42/ is actually an equality. This is also the case with other constructs seen so far since, as noted, they do not introduce imprecision.

Soundness proofs should similarly be provided for every instruction, although they do not appear in the present article.

## A.2 Object-oriented constructs

Adapting the previous formal model for the object-oriented version of the language, E2, involves changing the representation of states and the signature of instructions. The state now involves a set of objects, where each object may contain references to other objects. An instruction, previously a relation in $State \leftrightarrow State$, now has the signature $Object \rightarrow State \leftrightarrow State$; the use of an $Object$ as the first argument reflects the notion of current object and the principle of general relativity.

The full refinement of the formal model, and the corresponding proofs of soundness for the O-O rules of section 6 belong in another article.

## Appendix B: Computing the set of modified variables

The definition of the semantics of the alias calculus (3.3) uses $p\leftarrow$, the set of variables that every terminating execution of $p$ (a language construct) will set.

The computation of $p\leftarrow$ is undecidable. To prove this property (a case of Rice's theorem), we show that if we could compute $p\leftarrow$ we could also solve the halting problem. Consider the program

```
from over := False until over loop b end                    /43/
```

where $b$ is a terminating sequence of instructions (which may set the variable *over*). For any Turing-complete language, termination of /43/ for arbitrary $b$ is undecidable, even in the absence of any goto or other loop-exit instruction. Using a fresh variable *over1* we can rewrite the program, without changing its semantics, as

```
from
    over := False ; over1 := False
until over1 loop
    b
    if over then over1 := True end      <------  Loop body p
end
```

Calling the loop body ($b$ followed by the conditional) $p$, the loop terminates if and only if $over1 \in p\leftarrow$. So if we could compute $p$ we could decide the termination of /43/. ■

The impossibility of computing $p\leftarrow$ exactly in the general case is not a major problem in practice: as noted in 3.3, the Hoare-stye rule defining the semantics of the alias calculus remains valid, if no longer a weakest-precondition rule, if we replace $p\leftarrow$ by an under-approximation. The following rules define a sound under-approximation of $p\leftarrow$:

| | | |
|---|---|---|
| **skip**$\leftarrow$ | = | $\varnothing$ |
| (**create** $x$)$\leftarrow$ | = | $\{x\}$ |
| (**forget** $x$)$\leftarrow$ | = | $\{x\}$ |
| (**cut** $x, y$)$\leftarrow$ | = | $\{x, y\}$ |
| $(x := y)\leftarrow$ | = | $\{x\}$ |
| $(p\ ;\ q)\leftarrow$ | = | $p\leftarrow \cup\ q\leftarrow$ |
| (**then** $p$ **else** $q$ **end**)$\leftarrow$ | = | $p\leftarrow \cap\ q\leftarrow$ |
| $(p^n)\leftarrow$ | = | $p\leftarrow$ |
| (**loop** $p$ **end**)$\leftarrow$ | = | $\varnothing\leftarrow$ |
| (**call** $r$)$\leftarrow$ | = | $\underline{r}\leftarrow$ |
| (**call** $x.r$)$\leftarrow$ | = | $x\,\scriptstyle\blacksquare\,\underline{r}\leftarrow$ |

Under-approximation appears in the following cases: conditional (for **then** $p$ **else** $q$ **end** we only retain variables that appear in both $p\leftarrow$ and $q\leftarrow$ even if $q$, for example, is never executed); loop (where the result is $\varnothing$ to account for the case of zero executions, which might not occur); and qualified call (if $x$ is aliased to $y$ we could also include $y\,\scriptstyle\blacksquare\,\underline{r}\leftarrow$).

## Appendix C: Proof of the loop aliasing theorem

The loop aliasing theorem (4.10) states that for a given instruction $p$, if the sequence $t$ is defined by $t_0 = a$ and $t_{n+1} = t_n \cup (t_n \gg p)$, there exists an integer $N$ such that

1 For any $i < N$, $t_i \neq t_{i+1}$.

2 For any $i > N$, $t_i = t_N$.

3 $t_N = (a \gg \textbf{loop}\ p\ \textbf{end})$.

The first two properties are immediate:

- The sequence $t_n$ is non-decreasing over a finite set, and hence has a fixpoint.
- A non-decreasing sequence might encounter two or more equal consecutive elements (a plateau) before it reaches its fixpoint. This, however, cannot happen for a sequence defined in the form $t_{n+1} = f(t_n)$ (here $t_{n+1} = t_n \cup (t_n \gg p)$): if $t_N = t_{N+1}$, then $t_{N+2} = f(t_{N+1})$, also equal to $f(t_N)$ and hence to $t_{N+1}$ and $t_N$; all subsequent elements are equal as well. So the fixpoint is reached at the first $N$ such that $t_N = t_{N+1}$; this is the $N$ of the theorem.

Property 3 is informally clear if we consider **loop** $p$ **end** as equivalent to **skip** ; (**loop** $p$ **end** ; $p$), the fixpoint of the sequence $t_n$. For a more rigorous proof, let us show that $t_n$ is the same sequence as the sequence $s_n$ defined as

$$s_n \triangleq \bigcup_{i : 0 .. n} (a \gg p^n) \qquad /44/$$

This will give us the desired result since $a \gg$ **loop** $p$ **end**, defined in /25/ as $\bigcup_{n: \mathbb{N}} a \gg p^n$, is also as a consequence $\bigcup_{n: \mathbb{N}} s_n$; since $s_n \subseteq s_{n+1}$ for all $n$, the fixpoint of the sequence (the first $s_N$ such that $s_N = s_{N+1}$) will, if the sequences $s_n$ and $t_n$ are the same, yield $a \gg$ **loop** $p$ **end**.

The proof that the sequences are the same uses induction. First, $s_0 = t_0 = a$ and $s_1 = t_1 = (a \cup (a \gg p))$. (The induction step needs both base steps.) For the induction step, we prove separately that $s_{n+1} \subseteq t_{n+1}$ and that $t_{n+1} \subseteq s_{n+1}$. For the first property we expand the definition:

$$s_{n+1} = s_n \cup a \gg p^{n+1}$$

Since $s_n = t_n$ by the induction hypothesis and $t_n \subseteq t_{n+1}$ by the definition of $t$, it suffices to prove that $a \gg p^{n+1} \subseteq t_{n+1}$. By the definition of repetition, $a \gg p^{n+1} = (a \gg p^n) \gg p$. We note that $(a \gg p^n) \subseteq s_n$ by the definition of $s_n$ /44/, so $(a \gg p^{n+1}) \subseteq t_n$ by the induction hypothesis. This implies by monotonicity that $((a \gg p^n) \gg p) \subseteq (t_n \gg p)$ and hence (by the definition of the sequence $t_n$) that $((a \gg p^n) \gg p) \subseteq t_{n+1}$. This completes the proof that $s_{n+1} \subseteq t_{n+1}$.

For the induction step in the reverse direction, we expand the other definition :

$$\begin{aligned} t_{n+1} &= t_n \cup (t_n \gg p) && \text{-- By the definition of } t_n \\ &= s_n \cup (s_n \gg p) && \text{-- By the induction hypothesis} \end{aligned}$$

Since $s_n \subseteq s_{n+1}$ it suffices to prove that $(s_n \gg p) \subseteq s_{n+1}$. Since we have two base steps ($n = 0$ and $n = 1$), we may assume $n > 1$ and expand $s_n$ as $s_{n-1} \cup (a \gg p^n)$, so that by /27/ $s_n \gg p$ is $(s_{n-1} \gg p) \cup (a \gg p^{n+1})$; since the first operand is $t_{n-1} \gg p$ by the induction hypothesis and hence a subset of $t_n$ (which is also $s_n$), both terms are subsets of $s_{n+1}$. ■

## Appendix D: Towards a "must alias" calculus

The relation studied in this article describes when two variables *may* become aliased; it is usually an over-approximation of aliasings that do arise in program execution, first because not all executions will cause all possible aliasings, but also because the theory, as noted, may incur loss of precision. Some applications may need a "*must-alias*" theory, which can only err on the side of over-approximation.

While such a theory is beyond the scope of the present work, we may note that many of the rules of the may-alias calculus (collected in the next appendix) remain the same. The main difference will be that the rules for conditional /21/, repetition and loops /29/ will use intersection rather than union.

## Appendix E: The full calculus

Below for reference is the list of rules introduced for the calculus, preceded by the principal notations definitions on which they rely. The rule for the fixed repetition construct $p^n$ is omitted as this instruction is mostly useful as a stepping stone towards the loop construct.

| | | |
|---|---|---|
| $-$ | Set difference | |
| $a \setminus\!- E$ | $= a — \overline{A \times E}$ <br> -- i.e. $a$ deprived of all pairs <br> -- involving an element of $E$ | |
| $a / x$ | $= \{y\!: E \mid (y = x) \vee [x, y] \in a\}$ <br> -- i.e. all elements aliased to $x$ in $a$, plus <br> -- $x$ itself | |
| $a \gg$ **skip** | $= a$ | /12/ |
| $a \gg (p\ ;\ q)$ | $= (a \gg p) \gg q$ | /13/ |
| $a \gg$ (**forget** $x$) | $= a \setminus\!- \{x\}$ | /14/ |
| $a \gg$ (**create** $x$) | $= a \setminus\!- \{x\}$ | /15/ |
| $a \gg$ (**cut** $x, y$) | $= a — \overline{x, y}$ | /16/ |
| $a \gg (x := y)$ | $= a\,[x\!: y]$ | /17/ |
| $a\,[x\!: y]$ | $=$ **given** $b \triangleq a \setminus\!- \{x\}$ **then** <br> $\overline{b \;\cup\; (\{x\} \times (b / y))}$ <br> **end** | /18/ |
| $a\,[x\!: y]$ | $= (\ldots((a\,[x_1\!: y_1])\,[x_2\!: y_2])\ \ldots)\,[x_n\!: y_n]$ <br> -- For lists $x$ and $y$ | /20/ |
| $a$ » **then** $p$ **else** $q$ **end** | $= (a » p) \cup (a » q)$ | /21/ |
| $a$ » **loop** $p$ **end** | $= t_N$ <br> -- For the first $N$ such that $t_N = t_{N+1}$, <br> -- with $t_0 = a$ and $t_{n+1} = t_n \cup (t_n » p)$. | /29/ |
| $a » pr$ | $= (a » Main)$ <br> -- For a program $pr$ of main program $Main$ | /30/ |
| $a$ » **call** $r\ (l)$ | $= a\ [r^\bullet : l]\ »\ \underline{r}$ <br> -- Where is the list of formal arguments of $r$ | /31/ |
| **Current**.$e$ | $= e$ | /33/ |
| $e$.**Current** | $= e$ | /34/$x.x'$=**Current**/35/ |
| $x'.x$ | = **Current** | /36/ |
| $x.x'.e$ | $= e$ | /37/ |
| $x'.x.e$ | $= e$ | /38/ |
| $a$ » **call** $x.r\ (l)$ | $= x \blacksquare ((x' \blacksquare a)\ [r^\bullet : x' \blacksquare l]) » \underline{r}) \setminus\!- x \blacksquare r^\bullet$ | /40/ |
| $a$ » **call** $x.r$ | $= x \blacksquare ((x' \blacksquare a) » \underline{r})$ | /41/ |

## Appendix F: Acknowledgments

As noted in section 1, this work was made possible by the literature on software verification, particularly axiomatic semantics, separation logic, shape analysis, ownership types, dynamic frames and static analysis. The following references are available on these approaches:

- Axiomatic semantics: many references starting with C.A.R. Hoare, *An Axiomatic Basis for Computer Programming*, in *Communications of the ACM*, vol. 12, no. 10, Oct. 1969, pages 576–580. See the retrospective on this article in a recent issue of *Comm. ACM* at cacm.acm.org/magazines/2009/10/42360-retrospective-an-axiomatic-basis-for-computer-programming/fulltext.

- Spark: John Barnes, *High Integrity Software: The SPARK Approach to Safety and Security*. Addison-Wesley, 2003.

- Shape analysis: Mooly Sagiv, Thomas Reps and Reinhard Wilhelm, *Parametric shape analysis via 3-valued logic*, in *ACM Transactions on Programming Languages and Systems*, vol. 24, no. 3, May 2002, pages 217–298.

- Separation logic: many references starting with John C. Reynolds, *Separation Logic: A Logic for Shared Mutable Data Structures*, in *Logic in Computer Science*, 17th Annual IEEE Symposium, 2002, pages 55-74.

- Ownership types: David Clarke, John Potter and James Noble, *Ownership Types for Flexible Alias Protection*, in OOPSLA 1998, ACM SIGPLAN Notices, vol. 33, no. 10, Oct. 1998, pages 48-64.
- Dynamic frames: Ioannis Kassios, *Dynamic Frames: Support for Framing, Dependencies and Sharing Without Restrictions*, in *Formal Methods 2006*, eds. J. Misra, T. Nipkow and E. Sekerinski, Lecture Notes in Computer Science 4085, Springer Verlag, 2006, pages 268-283.
- Static analysis: see in particular Flemming Nielson, Hanne R. Nielson and Chris Hankin, *Principles of Program Analysis*, Springer Verlag, revised 2004.
- Spec# project at Microsoft Research: references available at research.microsoft.com/en-us/projects/specsharp/.
- JML (Java Modeling Language): references available at www.eecs.ucf.edu/~leavens/JML/.
- The frame problem: numerous references starting with Marvin Minsky, *A Framework for Representing Knowledge*, MIT-AI Memo 306, June 1974, also in *The Psychology of Computer Vision*, ed. P. Winston, McGraw-Hill, 1975. For applications to specification see Alex Borgida, John Mylopoulos and Raymond Reiter, *On the Frame Problem in Procedure Specifications*, IEEE Transactions on Software Engineering, vol. 21, no. 10, October 1995, pages 795-798.

John Hogg, Doug Lea, Alan Wills, Dennis deChampeaux and Richard Holt published in 1991 their "Geneva Convention on the Treatment of Object Aliasing", a general introduction (gee.cs.oswego.edu/dl/aliasing/aliasing.html) to aliasing issues in object-oriented languages.

This article has benefited from discussions with Scott West, Stephan van Staden, Carlo Furia, Cristiano Calcagno, Yi Wei, Alexander Kogtenkov and Sebastian Nanz. I am grateful to Peter O'Hearn, Reinhard Wilhelm and Tony Hoare for comments on the draft, and for comments on a related talk to Daniel Kröning (who gave me some advice towards ensuring modularity), Greg Nelson and Rick Hehner (who challenged me to show that no forward rule was possible, and found a problem with the initial definition of the ← operator).

I am particularly grateful to an anonymous referee appointed by the editor of this volume for finding a serious inconsistency in the original version of the semantics; his comments led to a significant improvement (section 3.3 and appendix B).

The continuing influence of Manfred Broy's work over many years, and the benefit of countless technical discussions with him, are gratefully acknowledged.

## Appendix G: References